\documentclass[12pt,preprint]{aastex}

\shorttitle{Pair analysis for Field Galaxies}
\shortauthors{Hsieh et al.}

\begin{document}

\title{Pair Analysis of Field Galaxies from the Red-Sequence Cluster Survey} 

\author{B. C. Hsieh\altaffilmark{1,6},
H. K. C. Yee\altaffilmark{2,6},
H. Lin\altaffilmark{3,6},
M. D. Gladders\altaffilmark{4,6},
D. G. Gilbank\altaffilmark{5}
}

\altaffiltext{1}{Institute of Astrophysics \& Astronomy, Academia Sinica,
P.O. Box 23-141, Taipei 106, Taiwan, R.O.C. Email: bchsieh@asiaa.sinica.edu.tw}

\altaffiltext{2}{Department of Astronomy \& Astrophysics, University of
Toronto, Toronto, Ontario, M5S 3H4, Canada. Email: hyee@astro.utoronto.ca}

\altaffiltext{3}{Fermi National Accelerator Laboratory, P.O. Box 500,
Batavia, IL 60510. Email: hlin@fnal.gov}

\altaffiltext{4}{Department of Astronomy and Astrophysics,
University of Chicago, 5640 South Ellis Avenue, Chicago, IL 60637, USA.
Email: gladders@oddjob.uchicago.edu}

\altaffiltext{5}{Astrophysics and Gravitation Group,
Department of Physics and Astronomy, University of Waterloo,
Waterloo, Ontario, N2L 3G1, Canada. Email: dgilbank@astro.uwaterloo.ca}

\altaffiltext{6}{Visiting Astronomer, Canada-France-Hawaii Telescope,
which is operated by the National Research Council of Canada, Le Centre
National de Recherche Scientifique, and the University of Hawaii.}

\begin{abstract}
We study the evolution of the number of close companions of similar
luminosities per galaxy ($N_c$) 
by choosing a volume-limited subset of the photometric redshift catalog 
from the Red-Sequence Cluster Survey (RCS-1). 
The sample contains over 157,000 objects with 
a moderate redshift range of $0.25 \leq z \leq 0.8$ and 
$M_{R_c} \leq -20$. 
This is the largest sample used for pair evolution analysis, 
providing data over 9 redshift bins with about 17,500 galaxies in each. 
After applying incompleteness and projection corrections, 
$N_c$ shows a clear evolution with redshift. 
The $N_c$ value for the whole sample grows with 
redshift as $(1+z)^m$, where $m = 2.83 \pm 0.33$
in good agreement with $N$-body simulations in a $\Lambda$CDM cosmology. 
We also separate the sample into two different absolute magnitude bins: 
$-25 \leq M_{R_c} \leq -21$ and $-21 < M_{R_c} \leq -20$,
and find that the brighter the absolute magnitude, the smaller the $m$ value.
Furthermore, we study the evolution of the pair fraction
for different projected separation bins and different luminosities.
We find that the $m$ value becomes smaller for larger separation,
and the pair fraction for the fainter luminosity bin has stronger evolution.
We derive the major merger remnant fraction $f_{rem} = 0.06$, 
which implies that about 6\% of galaxies with 
$-25 \leq M_{R_c} \leq -20$ have undergone major mergers since $z = 0.8$. 
\end{abstract}


\section{Introduction}\label{introduction}
Galaxy interactions and mergers play a very important role 
in the evolution and properties of galaxies. 
Although mergers are rare in the local universe,
a high merger rate in the past can change the morphology, luminosity, 
stellar population, and number density of galaxies dramatically. 
The evolution of the merger rate is directly connected to 
galaxy formation and structure formation in the Universe. 
Therefore, the measurement of merger rates for galaxies
at different merging stages provides important information
in the interpretation for various phenomena, such as galaxy and quasar
evolution.
There are many stages of mergers: 
close but well-separated galaxies (early-stage mergers), 
strongly interacting galaxies, 
and galaxies with double cores (late-stage mergers). 
Measuring the morphological distortion of galaxies
\citep[e.g.,][]{lefevre2000,reshetnikov2000,conselice2003,lavery2004,lotz2006},
as well as studying the spatially resolved dynamics of galaxies
\citep[e.g.,][]{puech2007a,puech2007b}
are definitive methods for studying on-going mergers,
However, they are challenging to observe at high redshift
and difficult to quantify.
Instead of directly studying on-going mergers,
close pairs are much easier to observe and
are still able to provide statistical information on the merger rate.
A close pair is defined as two galaxies which have 
a projected separation smaller than a certain distance.
Without redshift measurements, a close pair could be just an optical pair,
which contains two unrelated galaxies 
with small separation in angular projection.
With spectroscopic redshift measurements, 
a physical pair can be picked out
by choosing two galaxies with similar redshifts and small projected separation.
We note that only a fraction of physical pairs are real pairs,
which have true physical separations between galaxies 
smaller than the chosen separation,
and are going to merge is a relatively short timescale.
Nevertheless, studying any kind of close pairs allows us 
to glean statistical information on the  merger rate.

 From $\Lambda$ cold dark matter ($\Lambda$CDM) $N$-body simulations 
\citep{governato1999, gottlober2001}, 
it is found that the merger rates of halos increases with redshift 
as $(1+z)^m$, where $2.5 \leq m \leq 3.5$.  
However, observational pair results produce diverse results of
$0 \leq m \leq 4$ \citep[e.g.,][]{zepf1989, burkey1994, 
carlberg1994, woods1995, YE1995, patton1997, neuschaefer1997, lefevre2000, 
carlberg2000, patton2000, patton2002, bundy2004, lin2004, cassata2005,
bridge2007,kartaltepe2007}. 
Often, there are only a few hundred objects in these samples
because most studies use spectroscopic redshift data,
so that the error bars on the number of pairs are very large.
\citet{kartaltepe2007} utilized a photometric redshift database
and included 59,221 galaxies in their sample to perform the pair analysis.
The large sample makes their result robust.
However, a pair study with a 2 deg$^2$ field could still be affected
by cosmic variance ($\sim 30$ Mpc $\times$ 30 Mpc at $z = 0.5$).
Furthermore, some studies use morphological methods to 
identify pairs on optical images; 
some close pairs they find could just be late-type galaxies 
with starburst regions 
because they have double or triple cores and look asymmetrical. 
Therefore, the merger rates could be over-estimated. 
In contrast, other observations have yielded results 
with no strong evolution at high redshift 
\citep[e.g.,][]{bundy2004, lin2004}. 
\citet{bundy2004} also use a morphological method to identify pairs. 
However, they use near-IR images which reveal real stellar mass distributions 
and are insensitive to starburst regions. 
No matter what these previous studies conclude about pair fractions,
their results are obtained over a small number of redshift bins 
and have large error bars due to an inadequate number of objects
\citep[with the exception of][]{kartaltepe2007}.
Furthermore, none of these previous studies separate their sample
into field and cluster environments.
The evolution of the pair fraction can be very different 
in different environments.
The dynamics of galaxy in clusters is also very different 
compared to that in field;
hence, these studies could choose pairs 
with different properties even using exactly the same pair criteria.

In this paper, a subset of the photometric redshift catalog from
 \citet{hsieh2005}
is used to investigate the evolution of close galaxy pair fraction. 
The catalog is created using photometry in 
$z', R_c, V$, and $B$ from the Red-Sequence Cluster Survey 
\citep[RCS;][]{GY2005}. 
The sky coverage is approximately 33.6 deg$^2$. 
The large sample of photometric redshifts allows us 
to perform a pair analysis with good statistics. 
We use about 160,000 galaxies in our sample for the primary galaxies,
with a moderate redshift range of $0.25 \leq z \leq 0.8$ and 
$M_{R_c} \leq -20$,
allowing us to derive very robust results with relatively small error bars. 

This paper is structured as follows. 
In \S\ref{photoz} we briefly describe the RCS survey and 
the photometric redshift catalog used in our analysis. 
In \S\ref{sample} we provide a description of 
the sampling criteria for this study. 
Section \ref{method} presents the method of the pair analysis 
and the definition of a pair. 
We describe the selection effects and the methods 
dealing with the projection effects and incompleteness 
in \S\ref{selection_effect}. 
The error estimation of the pair fraction is discussed in \S\ref{err_est}. 
We then present the results in \S\ref{result} and 
discuss the implications of the pair statistics in our data 
in \S\ref{discussion}. 
In \S\ref{conclusion} we summarize our results and discuss future work. 
The cosmological parameters used in this study are 
$\Omega_\Lambda = 0.7$, $\Omega_M = 0.3$, $H_0 = 70$ km s$^{-1}$Mpc$^{-1}$, 
and $w = -1$. 

\section{The RCS Survey and Photometric Redshift Catalog}\label{photoz}
The RCS \citep{GY2005} is an imaging survey covering $\sim 92$ deg$^2$ 
in the $z'$ and $R_c$ bands carried out using the CFH12K CCD camera 
on the 3.6m Canada-France-Hawaii Telescope (CFHT) for the northern sky, 
and the Mosaic II camera on the Cerro Tololo Inter-American Observatory (CTIO) 
4m Blanco telescope for the southern sky, 
to search for galaxy clusters in the redshift range of $z < 1.4$. 
Follow-up observations in $V$ and $B$ were obtained 
using the CFH12K camera, covering 33.6 deg$^2$ 
($\sim$ 75\% complete for the original CFHT RCS fields).

The CFH12K camera is a 12k $\times$ 8k pixel$^2$ CCD mosaic camera, 
consisting of twelve 2k $\times$ 4k pixel$^2$ CCDs. 
It covers a 42 $\times$ 28 arcminute$^2$ area for the whole mosaic 
at prime focus (f/4.18), corresponding to 0".2059 pixel$^{-1}$. 
For the CFHT RCS runs, the typical seeing was 0.62 arcsec for $z'$ 
and 0.70 arcsec for $R_c$. 
The integration times were typically 1200s for $z'$ and 900s for $R_c$, 
with average $5\sigma$ limiting magnitudes of $z'_{AB}=23.9$ and $R_c=25.0$ 
(Vega) for point sources.
The observations, data, and the photometric techniques,
including object finding, photometric measurement,
and star-galaxy classification, are described in detail in \citet{GY2005}. 
For the follow-up CFHT RCS observations, the typical seeing was
0.65 arcsec for $V$ and 0.95 arcsec for $B$. 
The average exposure times for $V$ and $B$ were 480s and 840s, respectively, 
and the median $5\sigma$ limiting magnitudes (Vega) 
for point sources are 24.5 and 25.0, respectively.
The observations and data reduction techniques are described 
in detail in \citet{hsieh2005}. 

With the four-color ($z', R_c, V,$ and $B$) data, 
a multi-band RCS photometry catalog covering 33.6 deg$^2$ is generated.
Although the photometric calibration has been
done for all the filters \citep{GY2005,hsieh2005}, 
we refine the calibration procedure to achieve a better photometric
accuracy for this paper. First, we calibrate the patch-to-patch
zeropoints for $R_c$ by comparing the $R_c$ photometry of stars
to that in the SDSS DR5 database \footnote{\url{http://www.sdss.org/dr5/}}. 
We use the empirical transformation 
function between SDSS filter system and $R_c$ determined by Lupton (2005)
\footnote{\url{http://www.sdss.org/dr4/algorithms/sdssUBVRITransform.html}}.
The average photometry difference in each patch is the offset we apply.
However, patches 0351 and 2153 have no overlapping region with the SDSS DR5.
For these two patches,
we use the galaxy counts to match their zeropoints to the other patches.
The patch-to-patch zeropoint calibration of $z'$ is performed using
the same method.

We also utilize the SDSS DR5 database to calibrate the zeropoints
of $z'$ and $R_c$ between pointings within each patch. 
However, some patches do not entirely overlap with the SDSS DR5,
we have to use an alternative way \citep{glazebrook1994}
to calibrate the zeropoints for those pointings having no overlapping
region with the SDSS DR5.
There are fifteen pointings in each patch; the pointings overlap
with each other over a small area. 
By calculating the average differences of photometry 
for same objects in the overlapping areas for different pointings, 
we can derive the zeropoint offsets 
between pointings overlapping with the SDSS DR5 and 
those having no overlapping region with the SDSS DR5.
For patches 0351 and 2153, the pointing-to-pointing zeropoint
calibration is performed internally using the same method.
We note that we do not perform chip-to-chip zeropoint calibration 
for $z'$ and $R_c$ since it has been dealt with in great detail in
\citet{GY2005}.

For the $B$ and $V$ photometry,
we find by comparing to SDSS that there are systematic chip-to-chip 
zeropoint offsets in $B$, which are not found in $V$.
The transformation functions between SDSS $g, r, i$ 
and $V$, $B$ are from Lupton (2005).
In order to calibrate the zeropoints, we calculate the mean offset of each chip 
by combining all the chips with the same chip number, 
and then we apply the mean offsets to all the data in the corresponding chip.

Once the chip-to-chip calibration has been done,
the pointing-to-pointing and patch-to-patch zeropoint calibrations 
for $V$ and $B$ are performed using the same method as that for $z'$ and $R_c$.
For patches 0351 and 2153, which do not have SDSS overlap,
we utilize the stellar color distributions of 
$V-R_c$ and $B-R_c$ to obtain more consistent calibrations. 
Here, stars with $R_c < 22$ are selected to determine the magnitude offsets 
on a pointing-to-pointing and patch-to-patch basis. 
For the pointing-to-pointing calibration within a patch, 
all the stars with $R_c < 22$ in the patch 
are used as the comparison reference set.
The magnitude offsets in $V$ and $B$ for a pointing 
are computed using cross-correlation
between the reference data set and 
the data of the pointing of the stellar color distributions 
of $V$ - $R_c$ and $B$ - $R_c$, respectively.
The final pointing-to-pointing magnitude offsets 
are the summations of the offsets from the three iterations. 
The patch-to-patch photometry calibrations are performed 
after the pointing-to-pointing calibrations are done. 
In this case, all stars with $R_c < 22$ in all patches
are used as the comparison reference set.
The same calibration procedure as the pointing-to-pointing calibration 
is applied to the patch-to-patch calibration.

We note that object detection is performed by the Picture Processing Package
\citep[PPP,][]{yee1991} and has been found to reliably separate close pairs
over the separations of interest in this paper, and to not overly deblend
low-redshift galaxies to create false pairs.

\citet{hsieh2005} provides a photometric redshift catalog from 
the RCS for 1.3 million galaxies using 
an empirical second-order polynomial fitting technique 
with 4,924 spectroscopic redshifts in the training set. 
Since then there are more spectroscopic data available 
for the RCS fields and they can be added to the training set. 
The DEEP2 DR2 \footnote{\url{http://deep.berkeley.edu/DR2/}} overlaps 
with the RCS fields and provides 4,297 matched spectroscopic redshifts 
at $0.7 < z < 1.5$. 
The new training set not only contains almost twice the number of objects 
comparing to the old one but also provides a much larger sample for $z > 0.7$, 
i.e., better photometric redshift solutions for high redshift objects. 
Besides using a larger training set, 
we also use an empirical third-order polynomial fitting technique 
with 16 $kd$-tree cells to perform the photometric redshift estimation 
to achieve a higher redshift accuracy, 
giving an rms scatter $\sigma(\Delta{z}) < 0.05$ 
within the redshift range $0.2 < z < 0.5$
and $\sigma(\Delta{z}) < 0.09$ over the whole redshift range of $0.0 < z < 1.2$.
As in the original catalog, the new catalog also includes 
an accurately computed photometric redshift error 
for each individual galaxy determined by Monte-Carlo simulations 
and the bootstrap method, 
which provides better error estimates used for the subsequent science analyses. 
Detailed descriptions of the photometric redshift method 
are presented in \citet{hsieh2005}.

\section{Sample Selection}\label{sample}
\subsection{Estimating Absolute Magnitudes $M_{Rc}$}
In choosing a sample for an evolution study, great care must be taken. 
If the sampling criteria pick up objects with different physical properties 
(e.g., mass) at different epochs, the inferred evolution could be biased. 
For our pair study, we choose a volume-limited sample to 
include objects within the same range of absolute magnitude $M_{R_c}$ 
over the redshift range. 
The following equation is used for 
apparent magnitude to absolute magnitude conversion:
\begin{eqnarray}\label{absolutemag}
M_{R_c} = R_c - 5log{D_L} - 25 - k + Qz,
\end{eqnarray}
where $M_{R_c}$ is the absolute magnitude in $R_c$; 
$R_c$, the apparent magnitude of filter band $R_c$; 
$D_L$, the luminosity distance; $k$, the $k$-correction; 
$Q$, the luminosity evolution in magnitude; and $z$, the redshift. 

The traditional way of deriving the $k$-correction value for each galaxy is 
to compute its photometry by convolving an SED 
from a stellar synthesis model \citep{bc2003}
or an empirical template \citep{CWW1980} 
with the filter transformation functions. 
However, without spectroscopic redshift measurements, 
the accuracy of the $k$-correction value is affected by 
the less accurate photometric redshift. 
Without good $k$-correction estimations, 
the sample selection using absolute magnitude 
as one of the selecting criteria is problematic. 

\begin{figure}
\plotone{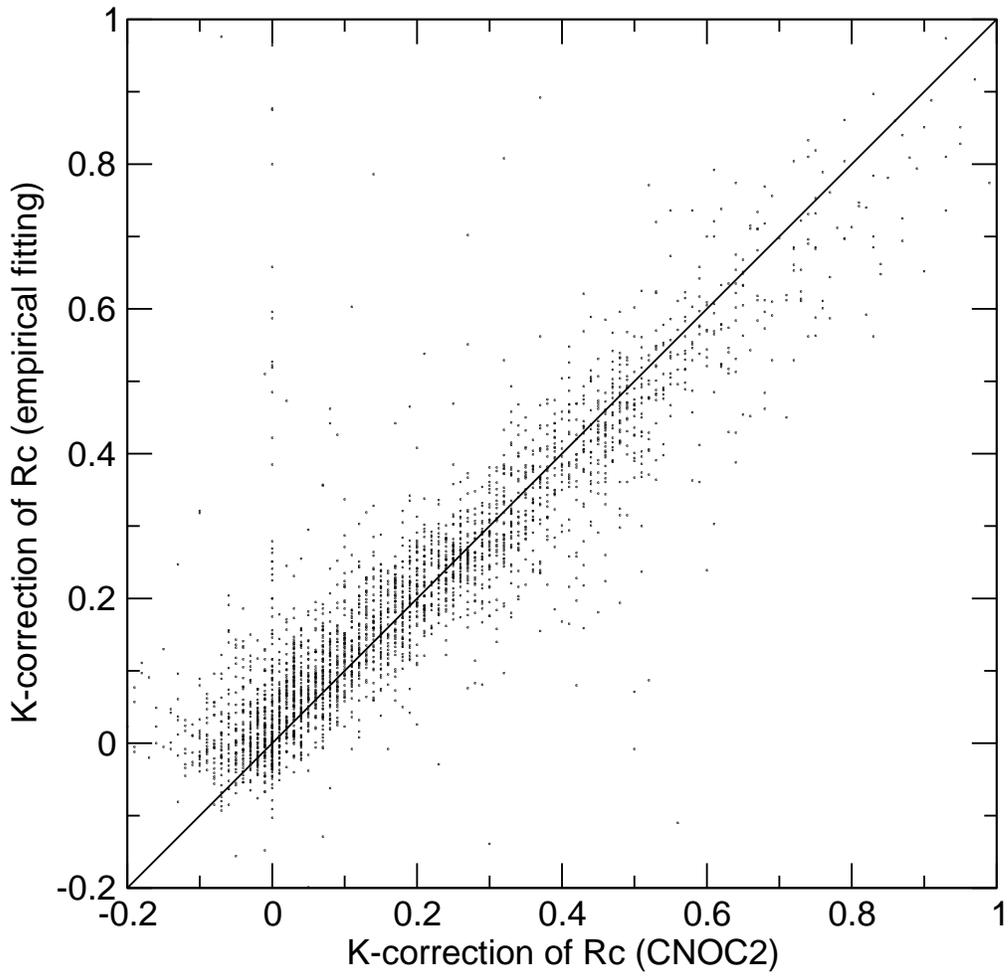}
\caption[The comparison between the empirical computed K-correction and 
the K-correction from the CNOC2 catalog]
{The comparison between the empirical computed K-correction and
the K-correction from the CNOC2 catalog. The rms scatter value is
about 0.03 mag.
\label{kcorr} }
\end{figure}

For our analysis, we use a new empirical method to 
determine the $k$-correction for the photometric redshift sample.
The Canadian Network for Observational Cosmology (CNOC2)
Field Galaxy Redshift Survey \citep{yee2000} is a survey 
with both spectroscopic and photometric measurements of galaxies, 
and the catalog provides $k$-correction values estimated by 
fitting the five broad-band photometry 
using the measured spectroscopic redshift
with empirical models derived from \citet{CWW1980}.
It allows us to generate a training set to determine the relation 
between $k$-correction and photometry, 
i.e., $k$-correction = $f$($z', R_c, V, B$).
We perform a second-order polynomial fitting of the $k$-corrections
for the training set galaxies and the result is shown in Figure~\ref{kcorr}. 
This plot shows that the $k$-correction can be estimated very well 
with rms scatter = 0.03 mag by this method;
we use the second-order polynomial fitting to 
derive the $k$-correction value for each galaxy.

We adopt $Q = 1.0$ for the luminosity evolution according to \citet{lin1999}. 

\subsection{Completeness}\label{completeness-correction}
The RCS photometric redshift catalog provides the 68\% 
($\sim1\sigma$ if the error distribution is Gaussian) 
computed photometric redshift error for each object. 
The photometric redshift uncertainty depends on 
the errors of the photometry for $z', R_c, V, B,$ and colors. 
The computed error is estimated using a combination of the bootstrap and 
Monte-Carlo methods and it has been confirmed empirically to be very reliable. 
The details are described in \citet{hsieh2005}.

For the sample used in the pair analysis, 
a photometric redshift error cut has to be defined. 
A looser error cut will make the sample more complete, 
but the final result will be noisier 
because data with larger errors are included. 
A tighter error cut will make the final result cleaner, 
but the incompleteness and bias problems will be more severe. 
For example, the bluer objects tend to have larger photometric redshift error 
and will be rejected more easily than red objects. 
If the pair result is color-dependent, 
the conclusion could be biased. 
As a compromise, we choose $\sigma_z/(1+z) \leq 0.3$ 
to be the criterion for the photometric redshift error cut for the sample. 
This criterion can eliminate those objects with catastrophic errors, 
and at the same time, 
the incompleteness and bias problems will not be too severe. 
Furthermore, we deal with the incompleteness and bias problems 
with completeness correction 
(see \S~\ref{completeness} for details) to minimize this selection effect.

\begin{figure}
\plotone{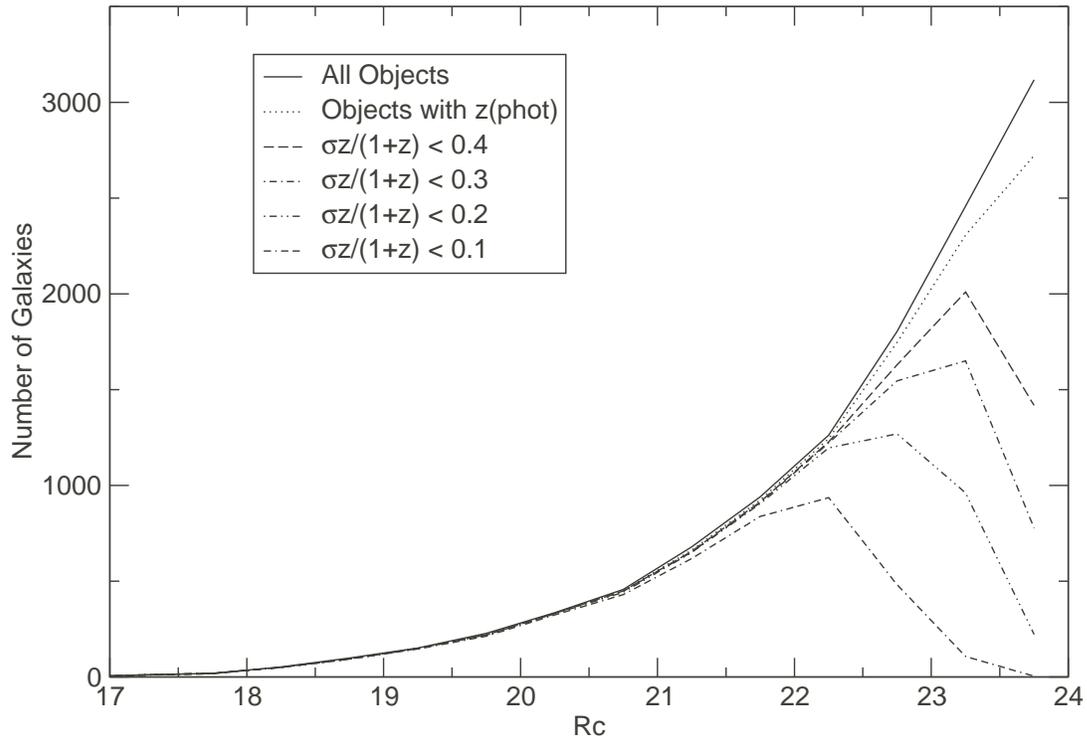}
\caption[$R_c$ histogram for pointing 0224A1 with different sampling criteria.]
{$R_c$ histogram for pointing 0224A1 with different sampling criteria.
The curves from top to bottom indicate all objects, 
objects with photometric redshift solutions, 
and those with $\sigma_z/(1+z) \leq 0.4$ to 0.1. 
It is clear to see that tighter criterion suffers from 
worse incompleteness.
\label{0223A1.complete} }
\end{figure}

The completeness correction factor is estimated
using the ratio of the total number of galaxies 
within a 0.1 mag bin at the $R_c$ magnitude of the companion 
to the number of galaxies in that bin 
with photometric redshifts satisfying the redshift uncertainty criterion.
The incompleteness problem is more severe with fainter magnitudes 
because of the larger photometry uncertainty. 
Figure~\ref{0223A1.complete} represents 
an example of $R_c$ galaxy count histograms
for subsamples of various $\sigma_z$ criteria for pointing 0224A1. 
The curves from top to bottom indicate the results with selecting 
criteria: all objects, objects with photometric redshift solutions, 
and those with 
$\sigma_z/(1+z) \leq 0.4$, $\sigma_z/(1+z) \leq 0.3$, 
$\sigma_z/(1+z) \leq 0.2$, and $\sigma_z/(1+z) \leq 0.1$. 
It is clear that tighter criteria suffer from worse incompleteness 
at fainter magnitudes.

\begin{figure}
\plotone{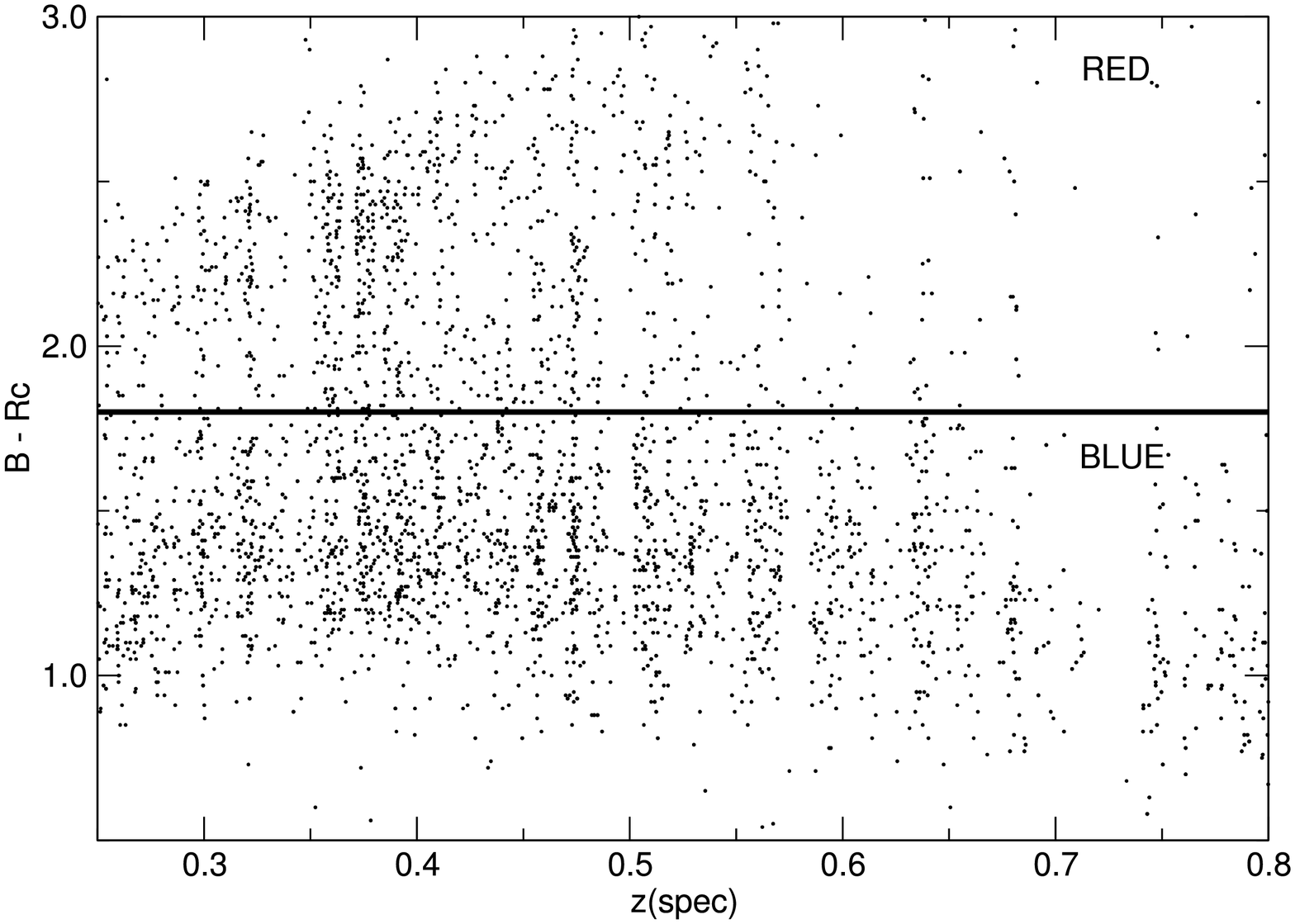}
\caption[Observed Color($B - R_c$) vs. spectral redshift from 
the training set of the RCS photometric redshift catalog.]
{Observed color($B - R_c$) vs. spectral redshift from the \emph{training set} 
of the RCS photometric redshift catalog \citep[see][for details]{hsieh2005}.
There are two loci in this plot. 
Most objects in the upper locus are early-type (red) galaxies, 
and for the lower locus, most of them are late-type (blue) galaxies. 
For the redshift range we study ($0.25 \leq z \leq 0.8$), 
a simple constant color cut at $B - R_c = 1.8$ 
can roughly separate the different types of galaxies.
\label{color-specz} }
\end{figure}

However, the completeness of the sample depends 
not only on magnitudes of $z', R_c, V,$ and $B$ 
but also on colors (e.g., galaxy types). 
At lower redshift, red objects are more complete than 
blue objects because early-type (red) galaxies 
have a clearer 4000\AA \ break which results in 
better photometric redshift fitting. 
At higher redshift, the incompleteness for red objects is getting worse 
(even worse than that of the blue objects) 
because the photometry of the blue filters 
is sufficiently deep for blue objects but not for red objects.
We refine our completeness correction by computing the factor separately
for red and blue galaxies. 
Figure~\ref{color-specz} represents the observed $B-R_c$ vs. spectral 
redshift relation. 
Most objects in the upper locus are early-type galaxies, 
and for the lower locus, most of them are late-type galaxies. 
For the redshift range of our sampling criterion ($0.25 \leq z \leq 0.8$), 
we can simply use $B - R_c = 1.8$ to 
roughly separate different types of galaxies. 

\subsection{Choosing the Volume Limit}\label{volume-limit}
There are two ways to select samples to perform a pair analysis. 
One is volume-limited selection, the other is flux-limited selection. 
The volume-limited selection is a proper way to pick up objects 
with the same characteristics. 
However, most previous pair studies select flux-limited samples 
and then try to correct these samples to volume-limited samples 
using Monte-Carlo simulations \citep[e.g.,][]{patton2000,patton2002,lin2004},
because of the small number of galaxies in their database. The RCS photometric 
redshift catalog contains more than one million objects; we can simply select 
a volume-limited sample and the number of objects is still statistically 
adequate for pair analysis.

\begin{figure*}
\plotone{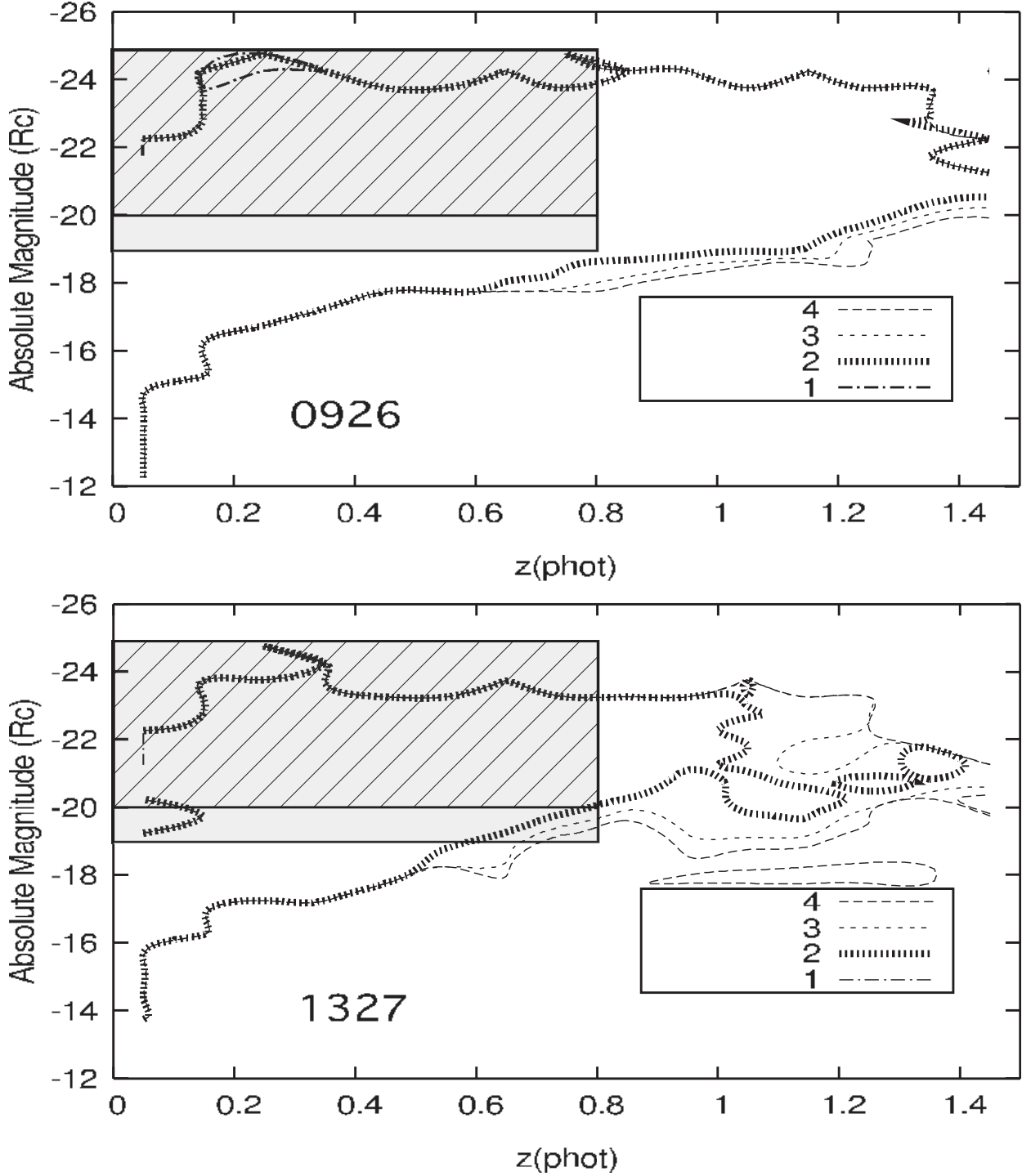}
\caption[Data completeness]
{Average data completeness in the absolute $R_c$ magnitude vs.
photometric redshift diagrams.
The upper panel is for patch 0926 and 
the lower panel is for patch 1327.
The contours indicate the completeness correction factors of
1, 2, 3, and 4.
According to these two panels, 
patch 0926 has much better completeness than patch 1327
in both the absolute $R_c$ magnitude axis and the photometric redshift axis.
We mark our limiting correction factor used to select our sample
by the heavy hashed lines.
The hatched region shows the selection area of our primary sample;
whereas the light shaded area, the secondary sample
(see \S\ref{boundary_effect} for details).
\label{absR-photoz} }
\end{figure*}

We choose our volume-limited by examining the relationship between
the completeness correction factor and absolute magnitude as
a function of redshift.
Figure~\ref{absR-photoz} shows the average data completeness in
the absolute $R_c$ magnitude vs. photometric redshift diagrams,
with a $\sigma_z/(1+z) \leq 0.3$ cut.
The upper panel is for patch 0926 and the lower panel is for patch 1327.
The contours indicate the completeness correction factors of
1, 2, 3, and 4.
According to Figure~\ref{absR-photoz}, 
patch 0926 has much better completeness than
patch 1327 in both the absolute $R_c$ magnitude axis and 
the photometric redshift axis.

Figure~\ref{absR-photoz} allows us to set reasonable ranges
in both the absolute $R_c$ magnitude axis and photometric redshift axis
for a volume-limited sample.
We decide to include data with completeness correction factors 
less than 2 in our sample, 
i.e., all the data in our volume-limited sample are at least 50\% complete
(marked by the heavy hashed lines).
While the sample is not 100\% complete, 
the completeness problem can be dealt with
the completeness correction later. (See \S\ref{completeness} for details.)
Hence, based on Figure~\ref{absR-photoz} 
and the diagrams for the other patches,
we choose a volume-limited sample using the following criteria: 
$0.25 \leq z \leq 0.8$, $-25 \leq M_{R_c} \leq -20$
(hatched region),
for which the completeness factor is less than 2.
We note that at $z=0.8$, and $M_{R_c} = -20$, 
early type galaxy has an $R_c \sim 24.7$.

\subsection{Choosing Field Galaxies}
The pair statistics could be very different in different environments. 
It is very interesting to study how the environment affects the pair result. 
However, for a cluster environment, 
the pair analysis is much more difficult and 
a significant amount of calibration/correction needs to be done 
because the pair signals could be embedded in a cluster environment and 
are difficult to delineate. 
We may need to develop a different technique for 
the pair analysis for the highest density regions (i.e., clusters). 
In this paper, we focus on the pair statistics in the field. 
The RCS cluster catalogs \citep{GY2005} are used 
to separate the field and the cluster regions. 
The cluster catalogs provide the information of 
the red-sequence photometric redshift, astrometry, and $r_{200}$, 
estimated using the richness $B_{gc}$ statistics of \citet{YE2003} in Mpc 
and in arcmin. 
An object inside $3 \times r_{200}$ of a cluster and 
having a photometric redshift within $z_{cluster} \pm 0.05$ 
is considered as a potential cluster member. 
Our rejection of clusters is fairly conservative;
we throw away pretty large regions 
in order to avoid biasing the field estimate,
at the expense of having a smaller field sample.
Approximately 33\% of the galaxies are rejected due to possibly
being in clusters.
The remaining objects are considered to be in the field
and included in our sample.

\section{Pair Fraction Measurement}\label{method}
We use the quantity $N_c$, 
defined as the number of close companions per galaxy, 
which is directly related to the galaxy merger rate, for our pair study. 
The definition of a close companion is a galaxy 
within a certain projected separation 
and velocity/redshift difference ($\Delta{z}$) of a primary galaxy.
After counting the number of close companions for each primary galaxy, 
we calculate the average number of close companions, which is $N_c$
(all the projected separations in this paper are in physical sizes,
unless noted otherwise).

Previous pair studies with spectroscopic redshift data 
\citep[e.g.,][]{patton2000,patton2002,lin2004} 
use pair definitions of 5-20 kpc to 5-100 kpc for projected separation, 
and a velocity difference $<$ 500km/s between 
the primary galaxy and its companions. 
However, because of the much larger redshift errors of 
the photometric redshift data, 
we cannot use the same criterion of $\Delta{v}$ as 
other spectroscopic pair studies for our analysis 
($\sigma_z = 0.05$ at $z = 0.5$ is equivalent to $\Delta{v} \sim 10000$km/s). 
Thus, to carry out the pair analysis using a photometric redshift catalog, 
we develop a new procedure which includes new pair criteria and 
several necessary corrections.

For the projected separation, 
we can use the same definition as the spectroscopic pair studies. 
For the redshift criterion, 
we utilize the photometric redshift error ($\sigma_z$) 
provided by the RCS photometric redshift catalog 
to develop a proper definition. 
We define the redshift criterion of a pair as $\Delta{z} \leq n\sigma_z$, 
where $\Delta{z}$ is the redshift difference 
between the primary galaxies and its companion, 
$n$ is a factor to be chosen, 
and $\sigma_z$ is the photometric redshift error of the primary galaxy. 
Due to the fact that the behavior of the pair fraction for major mergers 
(pairs with similar mass) and minor mergers 
(pairs with a huge difference of mass) could be different, 
one has to give a mass (luminosity) ratio limitation 
between the primary galaxy and its companion 
to specify the kind of merger being studied, i.e., 
$\Delta{R_c} \leq x$ mag.
We note that the absolute magnitude of $R_c$ should be used
for the luminosity/mass difference criterion.
However, since we use a photometric redshift catalog
to perform the pair analysis,
the redshifts of the primary and the secondary galaxies 
may be different due to the photometric redshift errors,
which would make the difference of the $k$-corrected absolute magnitudes
of the primary and the secondary galaxies 
larger than the luminosity difference criterion,
even if they actually fit the criterion.
Hence, we choose to use the apparent magnitude
instead of the absolute magnitude.

The quantity $N_c$ approximately equals to the pair fraction 
when there are few triplets or higher order $N$-tuples in the sample. 
In this study, $N_c$ will sometimes simply be referred to as the pair fraction.

\section{Accounting for Selection Effects}\label{selection_effect}
In this section, 
we discuss the effects of incompleteness, boundary, seeing
and projection, and the steps we take to account for them.
In all our subsequent discussions and analyses, we will use samples 
chosen with the following fiducial common criteria, unless noted otherwise:
$-25 \leq M_{R_c} \leq -20$ for the primary sample,
$-26 \leq M_{R_c} \leq -19$ for the secondary sample, with galaxies
selected satisfying $\sigma_z/(1+z) < 0.3$ and
$0.25 \leq z \leq 0.8$, and with the pair selection criteria of
$\Delta{z} \leq 2.5\sigma_z$,
5 kpc $\leq d_{sep} \leq$ 20 kpc, and $\Delta{R_c} \leq$ 1 mag.

\subsection{Completeness of the Volume-Limited Sample}\label{completeness}
For a pair analysis, 
the completeness of the sample is always an important issue. 
The pair fraction will drop due to the lower number density 
if the sample is not complete. 
Furthermore, the main point of this paper is to 
study the evolution of the pair fraction,
and unfortunately, the incompleteness is not independent of redshift: 
it gets more serious at higher redshifts. 
This effect is expected to make the pair fraction at higher redshifts
appear lower, and thus may bias conclusions regarding the redshift
evolution of the pair fraction.
Hence, to draw the correct conclusion, 
a completeness correction has to be applied.

\begin{figure}
\plotone{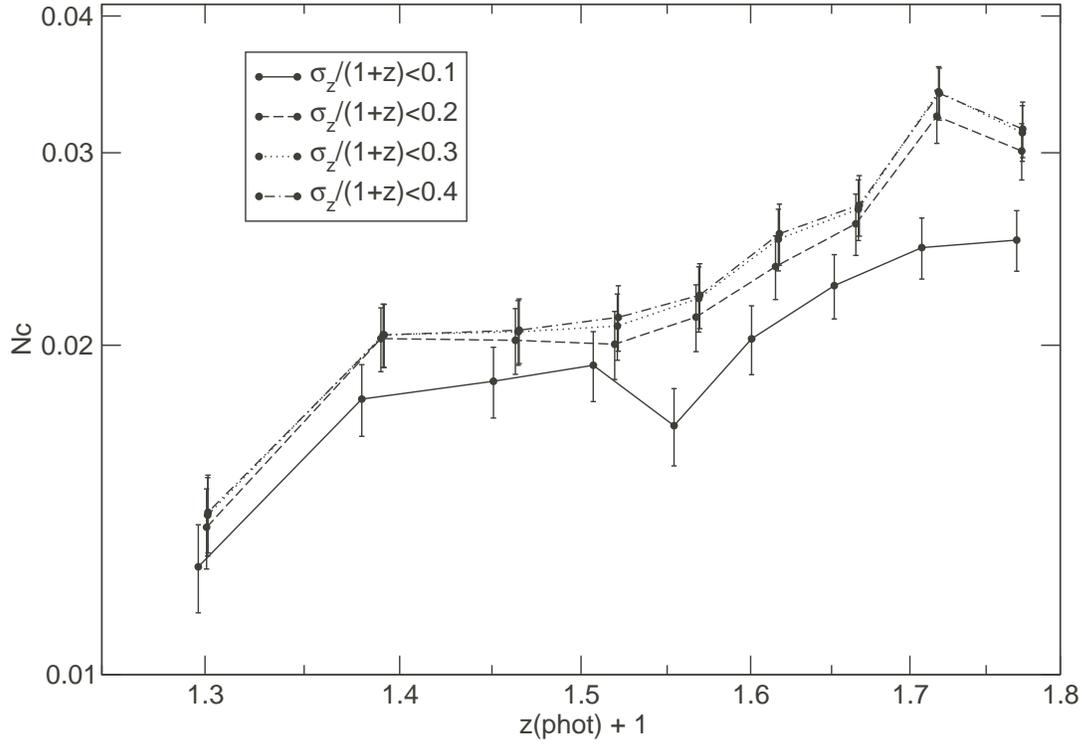}
\caption[Pair fractions with different $\sigma_z/(1+z)$ of 
the sampling criteria without the completeness correction 
(but with the projection correction).]
{Pair fractions with different $\sigma_z/(1+z)$ of 
the sampling criteria without the completeness correction 
(but with the projection correction). 
Each redshift bin contains from 15,500 objects
(for the $\sigma_z/(1+z) \leq 0.1$ criterion)
to 17,500 objects (for the $\sigma_z/(1+z) \leq 0.4$ criterion).
Note that the larger the $\sigma_z/(1+z)$, 
the higher the pair fraction, 
which is due to the lower completeness for samples with
a tighter redshift uncertainty criterion, causing some
companions to be missed in the pair counting.
\label{pair-zerr} }
\end{figure}

Figure~\ref{pair-zerr} illustrates the pair fractions 
for different sampling criteria in $\sigma_z/(1+z)$ 
without completeness corrections 
(but with projection correction, see \S\ref{backcor}).
Each redshift bin contains from 15,500 objects 
(for the $\sigma_z/(1+z) \leq 0.1$ criterion) 
The plot shows that the larger the $\sigma_z$ criterion, 
the higher the pair fraction. 
This is due to less completeness for a tighter criterion;
some real companions in pairs are missed 
because they have larger photometric redshift errors.
The differences between the pair fractions with 
different $\sigma_z/(1+z)$ criteria also become larger with redshift 
because the incompleteness is more severe at higher redshift,
especially for the $\sigma_z/(1+z) \leq 0.1$ criterion.
However, if a proper completeness correction is adopted, 
the pair results should not be affected by the sampling criteria 
and these curves of pair fraction with 
different $\sigma_z/(1+z)$ cuts should be similar.

We discussed the derivation of the completeness correction in
\S\ref{completeness-correction}, 
and after applying the completeness correction,
we find the pair fractions to be very similar 
for the different redshift uncertainty criteria,
which shows that the completeness correction works well.
The results after the completeness correction are shown in \S\ref{result}.

\subsection{Boundary Effects}\label{boundary_effect}
Objects near the boundaries of the selection criteria or 
close to the edges of the observed field could have fewer companions. 
This effect is referred to as a boundary effect. 
There are four different boundaries for our sample: 
$1)$ the boundaries of the selection criterion on the redshift axis;
$2)$ the edge of the observing field;
$3)$ the boundaries between the cluster region and the field;
and $4)$ the boundary of the absolute magnitude $M_{R_c}$ cut. 
The methods to deal with these boundary effects are 
described in this subsection.

The redshift range for our pair analysis is $0.25 \leq z \leq 0.8$. 
For primary galaxies at redshift close to 0.25 or 0.8, 
they will have fewer companions 
because some of their companions are scattered 
just outside the redshift boundaries and thus are not counted. 
To deal with this effect, 
we conduct the pair analysis for 
the full redshift range of the photometric redshift catalog ($0.0 < z < 1.5$), 
and then pick the primary objects 
satisfying the redshift criteria for the final result.

The objects near the boundaries of the observing field or 
near the gaps between CCD chips will have smaller pair fractions 
because some companions of these objects lie 
just across the boundaries of the observing field. 
The projected size of 20 kpc (the outer radius of pair searching circle) is 
about 5" (24 pixels) at redshift at 0.25; 
the size gets smaller at higher redshift. 
We limit the primary sample to be at least 5" from the edges of the CCDs 
while the secondary sample still includes all the objects. 
This configuration allows us to avoid the boundary effect 
of the limited survey fields.

Because we only study the pair fraction in the field for this investigation 
and separate our data into field environment and cluster environment, 
the boundary effect at the edges of $3 \times r_{200}$ and 
$z_{cluster} \pm 0.05$ is of concern. 
We constrain the primary sample to be at least $3 \times r_{200}$
from the centers of clusters and $\Delta{z} > 0.1$ from $z_{cluster}$. 
For the secondary sample, 
we limit it to be at least $3 \times r_{200} - 10"$ from 
the center of clusters and $\Delta{z} > 0.05$ from $z_{cluster}$. 
These new limits effectively eliminate these boundary effects.

The objects with absolute magnitudes close to 
the $M_{R_c}$ cut will have a lower pair fraction as well. 
This can be solved easily by searching for companions 
down to $M_{R_c} \leq -19$ for the primary sample with $M_{R_c} \leq -20$. 
Figure~\ref{absR-photoz} shows that the completeness correction factors
are still less than 2 
even when we push the $M_{R_c}$ boundary to -19 at $z \leq 0.8$ for patch 0926
(the light shaded area).
However, for patches 0351 and 1327,
the completeness is significantly worse than other 8 patches,
and they are substantially incomplete at $M_{R_c} = -19$.
Hence, they are abandoned in our pair analysis.
Consequently, the magnitude limit for the secondary sample is extended
to minimize this boundary effect.

\subsection{Projection Correction}\label{backcor}
Because of the projection effect due to 
the lower accuracy of the photometric redshifts
compared with spectroscopic redshifts,
a pair analysis using any criterion would find some false companions
and get a higher $N_c$ than the real value. 
To eliminate these foreground and background objects
from contaminating our results,
a projection correction is applied to our data.

\begin{figure}
\plotone{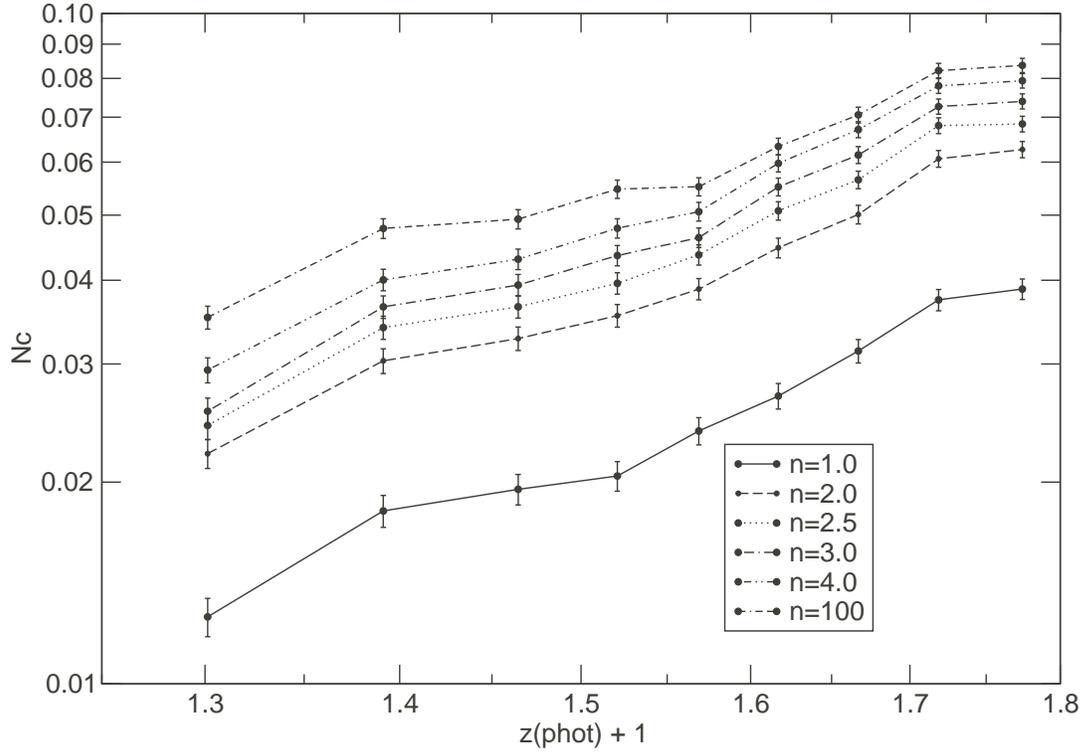}
\caption[Pair fractions with different $z_{primary} \pm n\sigma_z$ criteria
without any projection corrections (but with completeness corrections).]
{Pair fractions with different $z_{primary} \pm n\sigma_z$ criteria 
{\it without} any projection correction (but with completeness correction). 
Each data point contains about 17,500 objects. 
The pair fraction increases with larger $n$ value
because the pair criterion with larger $n$ includes 
more foreground/background galaxies and 
the results suffer from more serious projection effect. 
Note that the $n=100$ curve is approximately equivalent to not 
having photometric information.
\label{noback} }
\end{figure}

The magnitude of the projection effect on $N_c$ is illustrated in 
Figure~\ref{noback} which shows the results of using
the fiducial sample but with different
$z_{primary} \pm n\sigma_z$ criteria for inclusion of the companion without 
any projection correction (but with completeness correction).
Each data point contains about 17,500 objects. 
$N_c$ is higher with larger $n$ values
because the pair criterion with larger $n$ includes 
more foreground/background galaxies and 
the results suffer from more serious projection effect. 
For lower redshift, the projected search area is bigger than 
the one for higher redshift;
hence more foreground/background galaxies are counted. 
However, the photometric redshift error $\sigma_z$ 
becomes larger at higher redshift;
this effect also makes more foreground/background galaxies included. 
Therefore, from low redshift to high redshift, 
the projection effect affects the pair fraction about equally.

To correct for the projection effect, 
we calculate the mean surface density of all the objects in the same patch 
as the primary galaxy satisfying the pair criteria, 
$\Delta{z} \leq n\sigma_z$ and $\Delta{R_c} \leq 1$ mag, 
and then multiply it by the pair searching area (5-20 kpc) 
for each primary galaxy. 
This is the projection correction value for each primary galaxy.
The completeness corrections are also applied 
in the counting of the foreground/background galaxies. 
The final number of companions for each galaxy 
is the proper companion number 
with the projection correction value subtracted. 
Because not all primary objects have companions, 
some objects have negative companion numbers after the projection correction.

\subsection{The Effect of Seeing}\label{quality_effect}
The projected separation we use for the pair criteria is 
5 kpc $\leq d_{sep} \leq$ 20 kpc. 
For the highest redshift cut ($z = 0.8$) used in our analysis, 
the projected size is about three pixels (0".7) for 5 kpc (the inner radius). 
However, some data are taken in less favorable weather conditions, 
and the 5 kpc inner radius could be too small for these data 
due to poor resolution. 
In the meantime, we do not want to enlarge the inner radius 
because the closest pairs are the most important for a pair study. 
We have to make sure that 
the 5 kpc inner searching radius does not cause problems 
with data obtained in poorer seeing conditions. 

\begin{figure}
\plotone{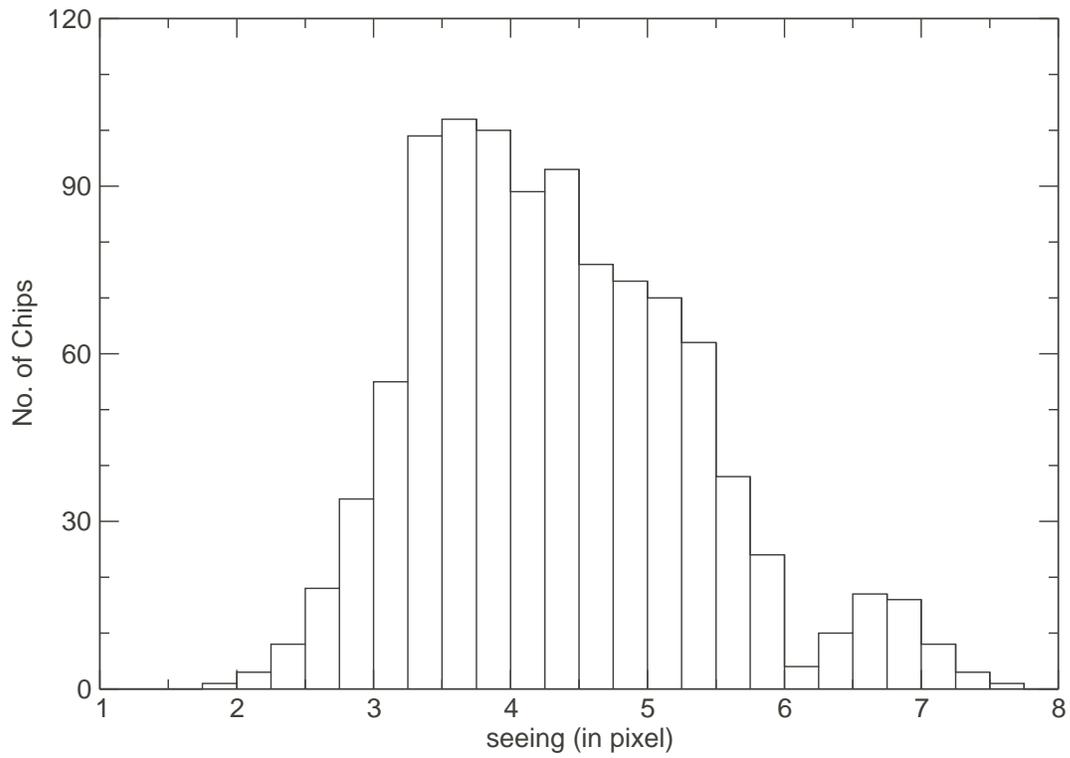}
\caption[Histogram of seeing condition for the $R_c$ images.]
{Histogram of seeing condition for the $R_c$ images. 
Most of the seeing values are about four pixels ($\sim 0".82$), 
but there are still some images taken under very poor weather 
with seeing up to 8 pixels.
\label{R-seeing} }
\end{figure}

\begin{figure}
\plotone{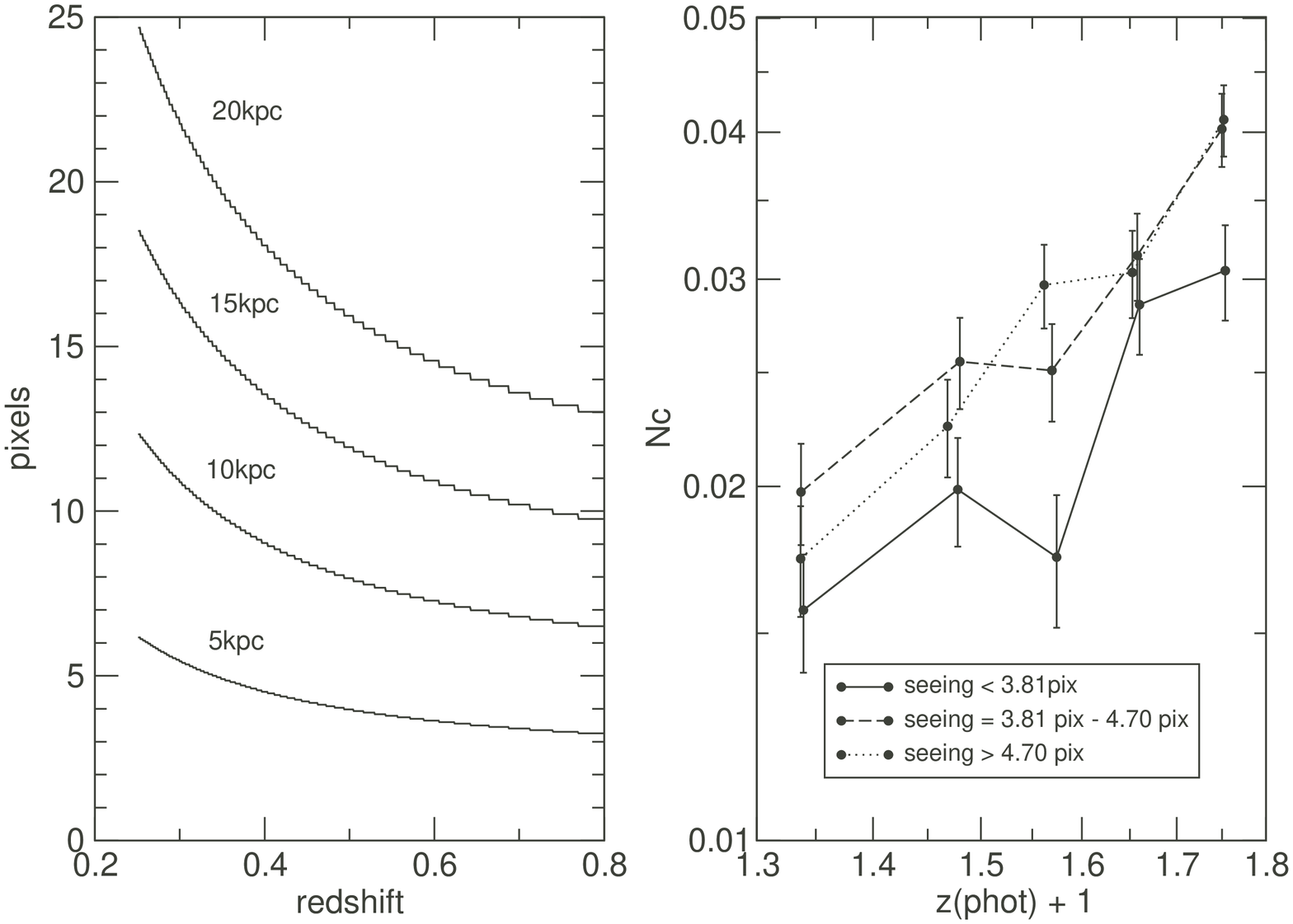}
\caption[Pair fractions of the data taken under different seeing conditions.]
{Pair fractions of the data taken under different seeing conditions.
The left panel represents the relation of the apparent size in pixels 
for different physical sizes (5, 10, 15, and 20 kpc) vs. redshift, 
as a reference, while the right panel shows the results 
with different seeing conditions. 
The curves in the right panel indicate the pair fraction of data 
taken under different weather conditions. 
There are about 10,000 objects in each redshift bin. 
We find that the curve with the best seeing condition is in fact
about 20\% lower than the other two worse seeing conditions,
indicating that the range of seeing in our data do not cause a drop in $N_c$.
We note that most pointings with the best seeing condition are 
from two patches;
the somewhat lower $N_c$ for the good seeing data could be 
just due to cosmic variance.
\label{seeing} }
\end{figure}

We only focus on the seeing conditions of the $R_c$ images 
because object finding was carried out 
using the $R_c$ images (see \citealt{hsieh2005} for details). 
The distribution of the seeing conditions of $R_c$ for the 8 patches 
is shown in Figure~\ref{R-seeing}. 
From this plot, most seeing values are between three to six pixels. 
To test if the inner radius (5 kpc) of the searching criterion 
is too small for data taken in poorer seeing conditions, 
we perform the pair analyses for three subsamples 
with different seeing conditions separately, 
separated at seeing $= 3.81$ pixels (0$"$.78) and $4.70$ pixels (0$"$.97). 
Each subsample contains a similar number of pointings. 
The results are shown in Figure~\ref{seeing}. 
The left panel represents the relation of the apparent size in pixels 
for different physical sizes (5, 10, 15, and 20 kpc) vs. redshift, 
as a reference, 
while the right panel shows the results 
with different seeing conditions. 
There are about 10,000 objects in each redshift bin. 
We find that the curve with the best seeing condition is in fact
about 20\% lower than the other two poorer seeing conditions,
indicating that the range of seeing in our data do not cause a drop in $N_c$.
We note that most pointings with the best seeing conditions are 
from patches 1416 and 1616;
the 20\% lower $N_c$ for the good seeing data could be 
just due to cosmic variance.
Therefore, the 5 kpc inner searching radius for all the data does not appear 
to produce a significant selection effect on the result.

\section{Error Estimation of $N_c$}\label{err_est}
The error of $N_c$ is estimated assuming Poisson distribution
and derived using:
\begin{eqnarray}\label{error}
error = \frac{\sqrt{N_{companion}+N_{proj}}}{N_{primary}}
\end{eqnarray}
where $N_{companion}$ is the total number of the companions,
$N_{proj}$ is the sum of the projection correction values,
and $N_{primary}$ is the number of the primary galaxies in each redshift bin.
We note that the values of $N_{companion}$ and $N_{proj}$
are the ones without the completeness corrections 
in order to retain the correct Poisson statistics.

\section{Results}\label{result}
\begin{figure}
\plotone{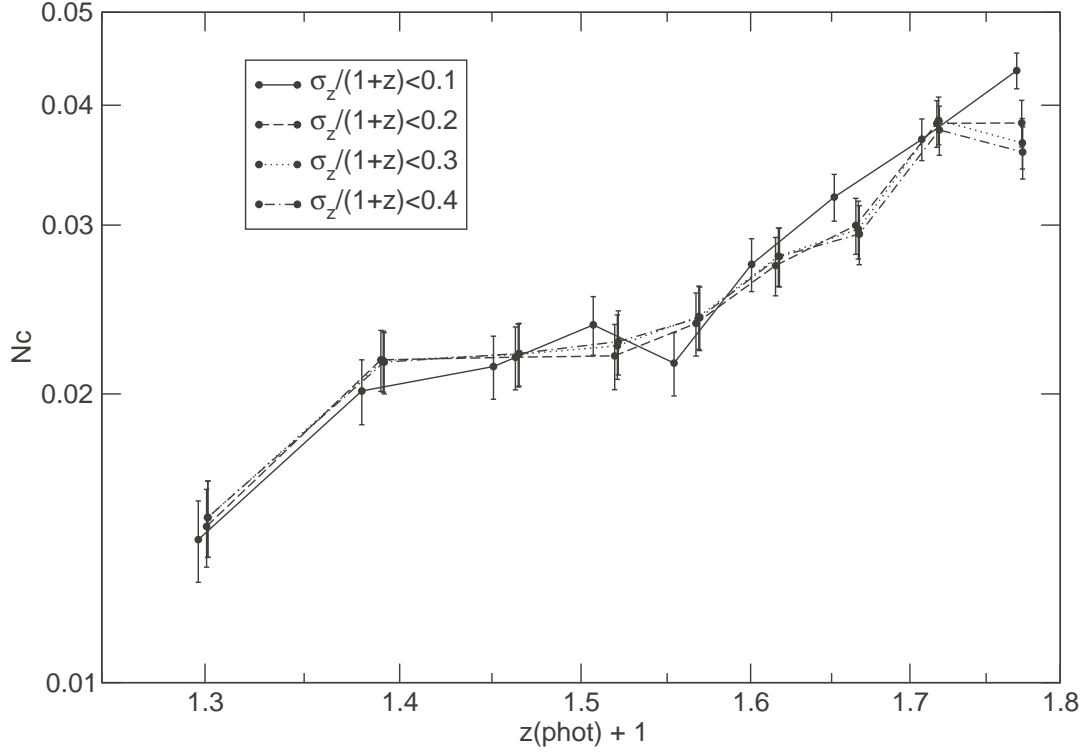}
\caption[Pair fraction vs. redshift after the completeness correction.]
{Pair fraction vs. redshift after the completeness correction, with
different redshift uncertainty (in $\sigma_z/(1+z)$) criteria, using the 
the sampling criteria: $-25 \leq M_{R_c} \leq -20$ for the primary sample,
$-26 \leq M_{R_c} \leq -19$ for the secondary sample, $0.25 \leq z \leq 0.8$,
and the pair criteria: 5 kpc $\leq d_{sep} \leq$ 20 kpc,
$\Delta{z} \leq 2.5\sigma_z$, $\Delta{R_c} \leq 1$ mag.
The completeness and projection corrections are applied.
Each redshift bin contains from 15,500 objects
(for the $\sigma_z/(1+z) \leq 0.1$ criterion)
to 17,500 objects (for the $\sigma_z/(1+z) \leq 0.4$ criterion).
Comparing to Figure~\ref{pair-zerr},
the curves of pair fractions are very similar 
for the different redshift uncertainty criteria,
which shows that the completeness correction works well.
\label{sigma_z} }
\end{figure}

\begin{figure}
\plotone{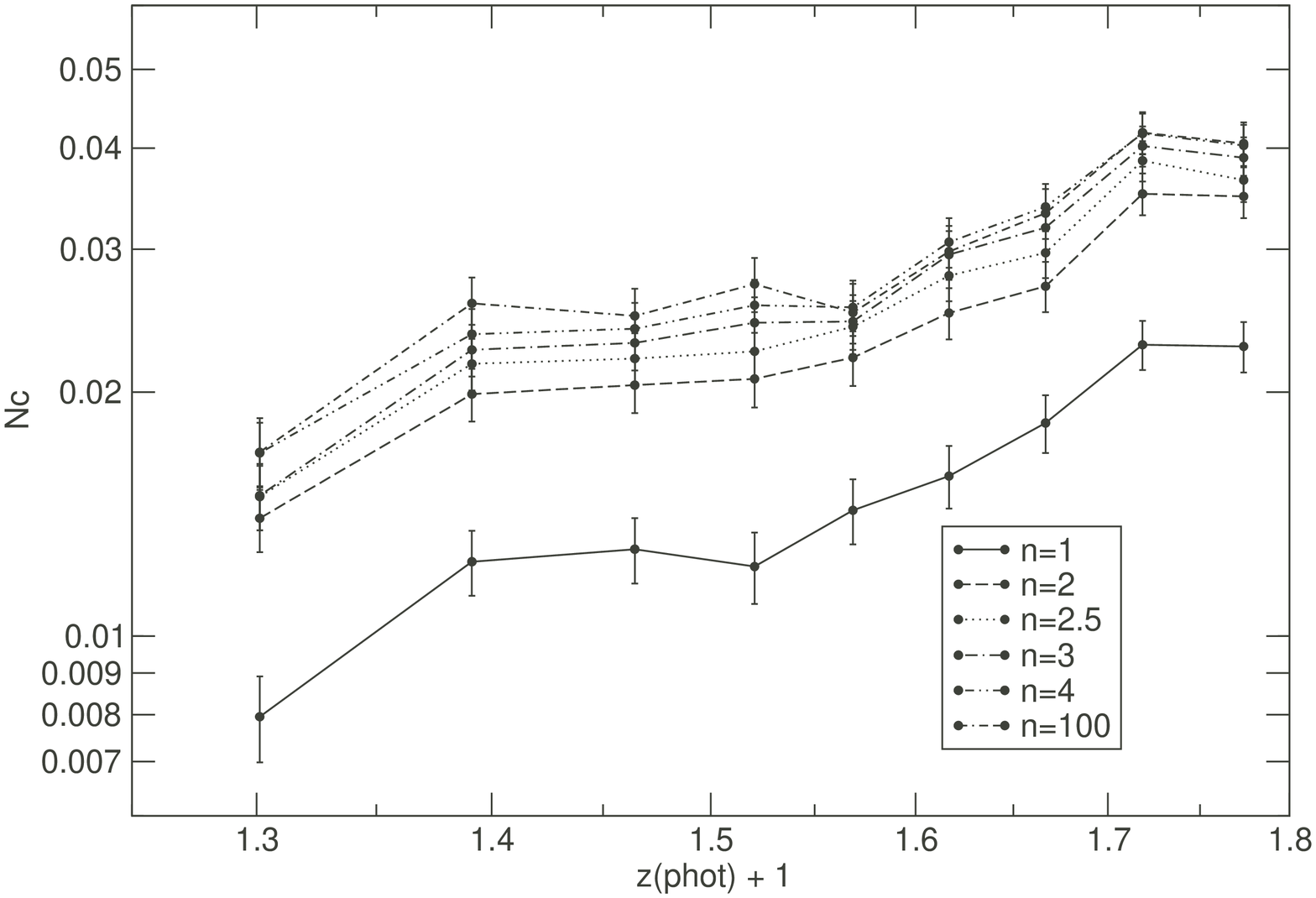}
\caption[Pair fraction vs. photometric redshift curves with different
$\Delta{z} = n\sigma_z$ criteria.]
{Pair fraction vs. photometric redshift curves with different 
$\Delta{z} = n\sigma_z$ criteria. 
The projected separation criterion used in this plot is 5-20 kpc. 
The completeness and projection corrections are applied. 
Pair fractions using the parameter 
$n = 1.0, 2.0, 2.5, 3.0, 4.0,$and $100.0$ are plotted. 
We use $n = 100.0$ to approximate $n = \infty$ 
which indicates the result using no redshift information for companions. 
Each data point contains 17,500 objects.
Note that curves with $n \leq 2.5$ are consistent with being the same.
The curve with $n = 1.0$ is expected to be about 32\% lower 
than the one with $n = 100.0$ if the distribution of
the photometric redshift error is Gaussian.
The $n = 1.0$ result is about 40\% and 30\% lower 
than the ones with $n = 100.0$, and $n = 2.5$, respectively.
\label{nX} }
\end{figure}

\begin{figure}
\plotone{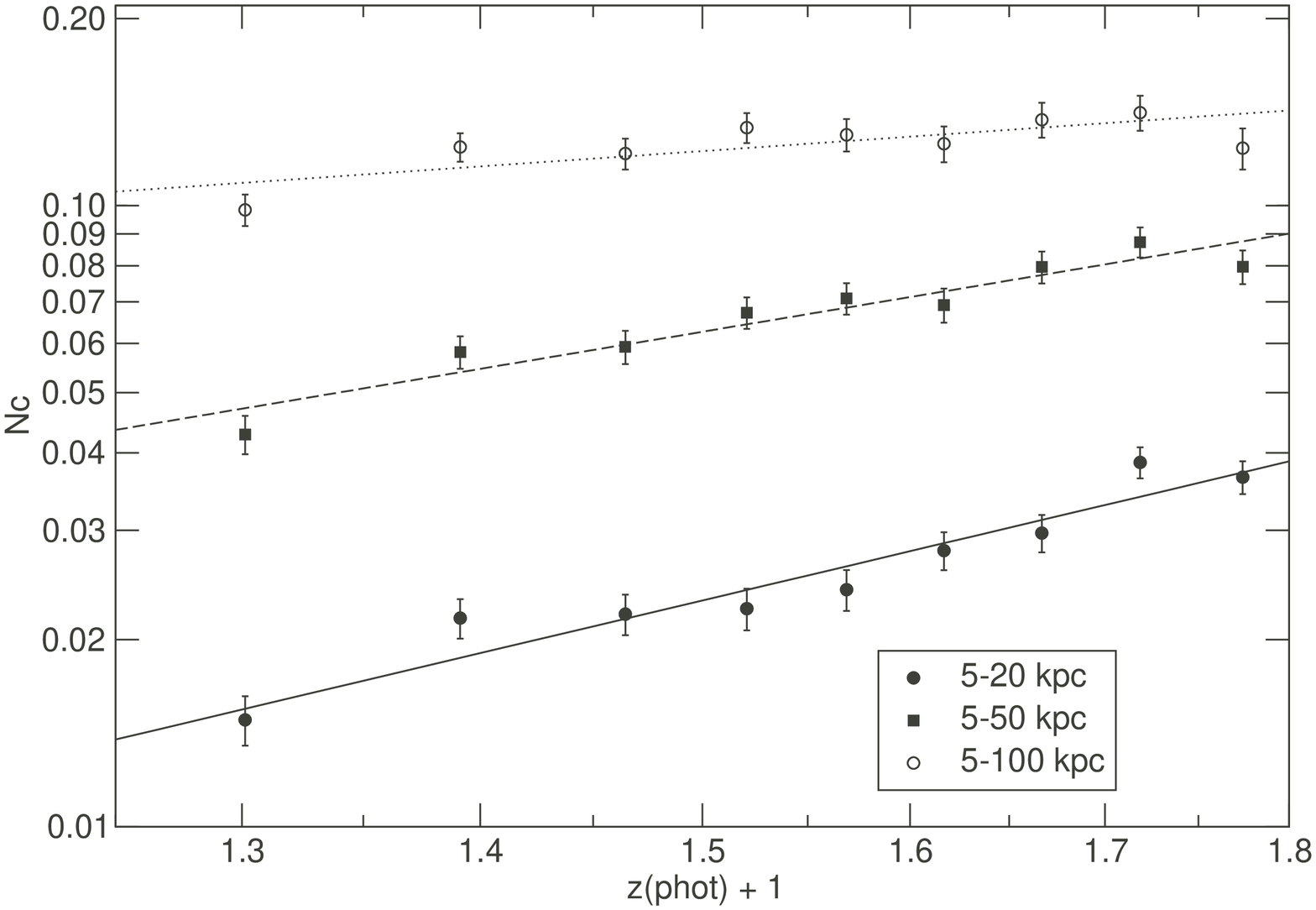}
\caption[Pair fractions using different projected separations for 
the pair criterion.]
{Pair fractions using different projected separations for the pair criterion. 
There are about 17,500 objects in each redshift bin. 
The completeness and projection corrections are applied. 
The inner radius is 5 kpc. 
Different line types and symbols indicate different outer radii 
from 20 kpc to 100 kpc. 
Generally speaking, the larger the outer radius, 
the higher the pair fraction because more galaxies satisfy the pair criteria.
\label{pair_sep} }
\end{figure}

\begin{figure}
\plotone{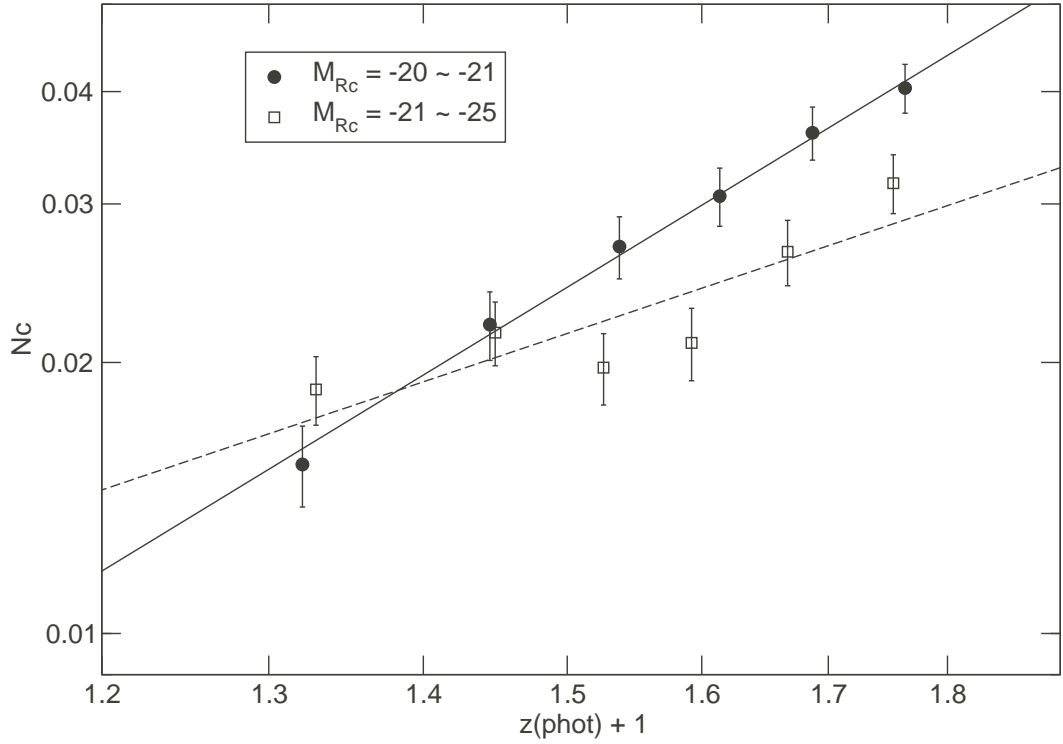}
\caption[Pair fractions with different $M_{R_c}$ bins.]
{Pair fractions with different $M_{R_c}$ bins. 
The filled circles and the open squares indicate
the absolute magnitude bins of $-21 \leq M_{R_c} \leq -20$ and
$-25 \leq M_{R_c} < -21$, respectively. 
The numbers of objects in each redshift bin with absolute magnitude cuts 
for faint and bright are 15,700 and 10,500, respectively.
We find that the pair fraction increases with
redshift as $(1+z)^m$ where $m = 3.25 \pm 0.11$ (solid line)
and $m = 1.79 \pm 0.53$ (dashed line)
for the absolute magnitude bins of
$-21 \leq M_{R_c} \leq -20$ and $-25 \leq M_{R_c} < -21$, respectively,
i.e., the brighter the absolute magnitude, the smaller the $m$ value.
\label{M_Rc} }
\end{figure}

\begin{figure}
\plotone{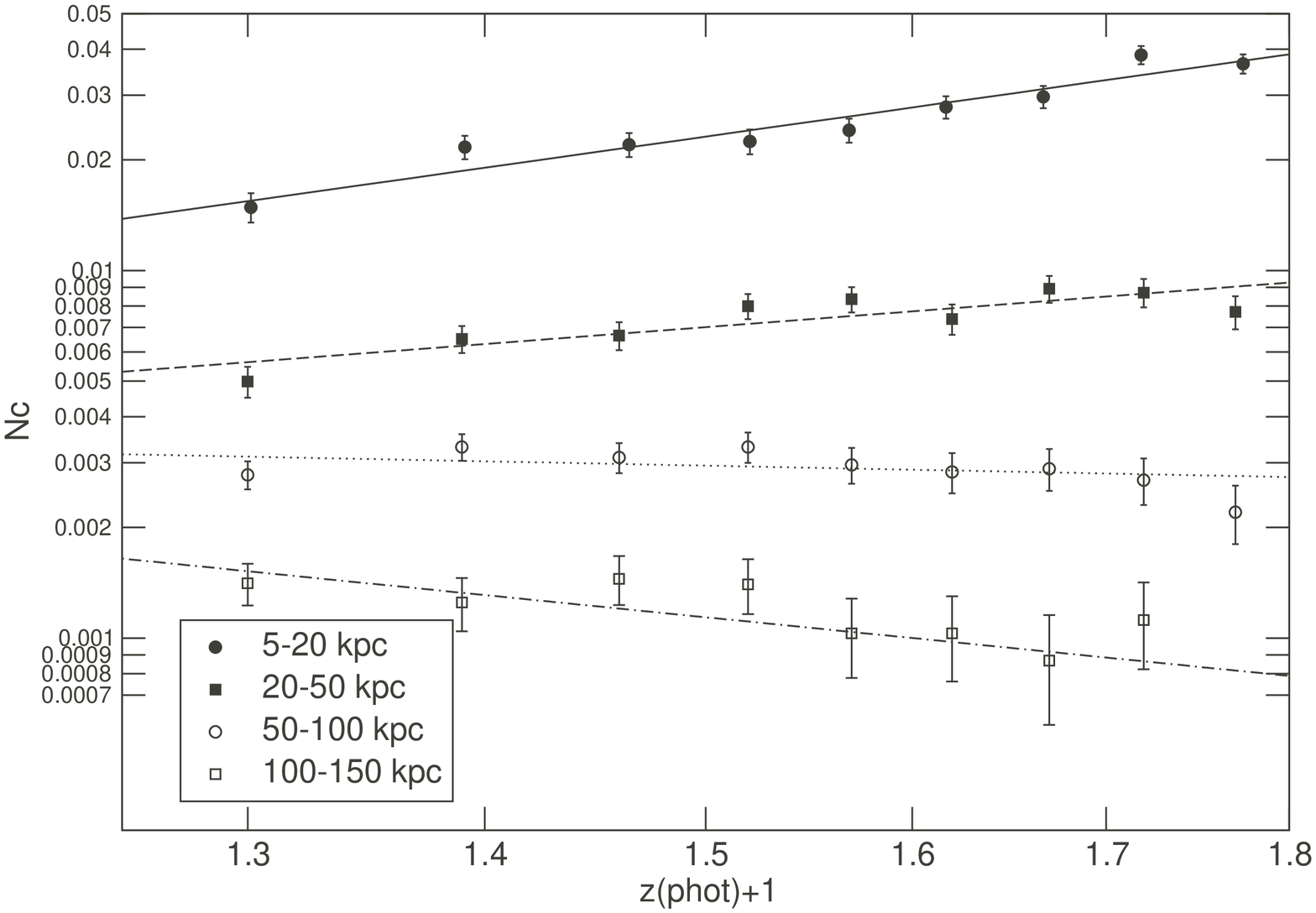}
\caption[Pair fractions with different projected pair separation bins.]
{Pair fractions with different projected pair separation bins. 
The number of objects in each redshift bin is about 17,500.
The evolutions of the pair fraction follow $(1+z)^m$
where $m = 2.83 \pm 0.33, 1.53 \pm 0.36, -0.39 \pm 0.36$,
and $-1.20 \pm 0.48$
for separation 5-20 kpc, 20-50 kpc, 50-100 kpc, and 100-150 kpc, respectively.
We note that the $m$ value decreases with increasing separation.
\label{sep-1} }
\end{figure}

The results of the pair analysis are shown 
in Figures~\ref{sigma_z} to \ref{sep-1}.
Because we focus only on major mergers, 
the difference in $R_c$ between the galaxies in close pairs
is restricted to be less than one magnitude. 
If the mass-to-light ratio is assumed to be constant 
for different types of galaxies at the same redshift, 
the mass ratio between the primary galaxies 
and companions ranges from 1:1 to 3:1. 

Figure~\ref{sigma_z} shows the results with different $\sigma_z/(1+z)$ cuts,
with the fiducial sampling criteria listed in \S\ref{selection_effect},
with  completeness and projection corrections applied.
The redshift bins contain from 15,500 objects 
(for the $\sigma_z/(1+z) \leq 0.1$ criterion)
to 17,500 objects (for the $\sigma_z/(1+z) \leq 0.4$ criterion).
Compared to Figure~\ref{pair-zerr}, 
the curves of the pair fractions are very similar for 
the different redshift uncertainty criteria,
which illustrates that the completeness correction works well.

For the redshift criterion of the pair definition, 
we count companions within $z_{primary} \pm n\sigma_z$, 
where $z_{primary}$ is the redshift of the primary galaxy
and $\Delta{z} = n\sigma_z$ (see \S\ref{method}).
If the photometric redshift error is assumed to be Gaussian, 
when $n = 1.0$, we count only about 68\% of the companions; 
when $n = 2.0$, about 95\%; and so on. 
Of course, the criteria with larger $n$ will include 
more foreground/background galaxies and make $N_c$ higher, 
but this can be dealt with by the projection correction 
(see \S~\ref{backcor} for details). 

Figure~\ref{nX} represents $N_c$ vs. photometric redshift curves 
using different $\Delta{z} = n\sigma_z$ criteria 
with the fiducial sampling criteria, with
completeness and projection corrections applied.
Pair fractions using the values $n = 1.0, 2.0, 2.5, 3.0, 4.0,$ 
and $100.0$ are plotted. 
We use $n = 100.0$ to approximate $n = \infty$ 
which is equivalent to the result using no redshift information for companions. 
Each data point contains about 17,500 objects. 
Curves with $n \geq 2.5$ are consistent with being the same.
As discussed, the curve with $n = 1.0$ should be about 32\% lower than 
the one with $n = 100.0$ if the distribution of 
the photometric redshift error is Gaussian. 
The curve with $n = 1.0$ is about 40\% lower than 
the one with $n = 100.0$, and about 30\% lower than the one with $n = 2.5$.
While the results do not perfectly match what is expected
from a Gaussian distribution of photometric redshift uncertainties,
comparing Figure~\ref{nX} to Figure~\ref{noback}, the application of
the projection correction does reduce the $N_c$ derived by approximately
the correct amount.
The somewhat larger difference is not unexpected, 
since the true probability distribution of the
uncertainty of the photometric redshift is likely non-Gaussian, but
somewhat broader.
We note that the error bars with larger $n$ are bigger 
because of larger projection correction errors
(see \S\ref{err_est} for details). 
We choose $n = 2.5$ for our final result;
with this, about 99\% of the companions are counted, 
which is a compromise between the completeness of the companion counting and 
the size of the error of the projection correction.

In Figure~\ref{pair_sep}, we show $N_c$ using different outer radii 
of the projected separations for the pair criterion. 
The completeness and projection corrections are applied. 
We use an inner radius of 5 kpc for all cases.
Different line types and symbols indicate different outer radii 
from 20 kpc to 100 kpc. 
As expected, the larger the outer radius, 
the higher the $N_c$ because more galaxies satisfy the pair criteria. 
For better statistics,
and because most previous pair studies use 5-20 kpc as their pair criteria
(since pairs with separations of 20 kpc will be almost certain to merge), 
we choose 5 kpc $\leq d_{sep} \leq$ 20 kpc to 
be the pair criterion for the projected separation for this study. 

Using 5 kpc $\leq d_{sep} \leq$ 20 kpc and $\Delta{z} \leq 2.5\sigma_z$, 
as our fiducial criteria,
we find that the pair fraction increases with 
redshift as $(1 + z)^m$ where $m = 2.83 \pm 0.33$.
The best fit is plotted as a solid line on Figure 11.
The error is estimated using the Jackknife technique.
We note that the error bar of the $m$ value is affected
not only by the error bar of each redshift bin
but also by whether the function $(1 + z)^m$ 
is a good representation of the data.

We also study the pair fractions with 
different absolute magnitude $M_{R_c}$ cuts and 
show the results in Figure~\ref{M_Rc}. 
The filled circles and the open squares indicate
the absolute magnitude bins of 
$-21 \leq M_{R_c} \leq -20$ and $-25 \leq M_{R_c} < -21$, respectively. 
The numbers of objects in each redshift bin 
with absolute magnitude cuts for faint and bright 
are about 15,700 and 10,500, respectively.
We find that the pair fraction increases with
redshift as $(1+z)^m$ where $m = 3.25 \pm 0.11$ (solid line)
and $m = 1.79 \pm 0.53$ (dashed line)
for the absolute magnitude bins of
$-21 \leq M_{R_c} \leq -20$ and $-25 \leq M_{R_c} < -21$, respectively;
i.e., the brighter the absolute magnitude, the smaller the $m$ value.
However, the error bar of the $m$ value is significantly larger for the
high-luminosity sample.
This effect is not due to the slightly larger error bars in the
bright sample, but rather that the $(1+z)^m$ power-law does not
appear to be a good representation
of the evolution of $N_c$ for the higher luminosity galaxies.
This result is further discussed in \S\ref{discussion}.

Based on Figure~\ref{pair_sep}, 
it is apparent that the larger the outer radii, the lower the $m$ value is.
To look at this in more detail,
we use rings of area for the pair counting,
and show the results of the evolution of $N_c$ in Figure~\ref{sep-1}.
We note that all $N_c$ values are normalized by area 
using the 5-20 kpc bin as the reference.
The number of objects in each redshift bin is about 17,500.
The evolution of the pair fraction follow $(1+z)^m$
where $m = 2.83 \pm 0.33, 1.53 \pm 0.36, -0.39 \pm 0.36$,
and $-1.20 \pm 0.48$
for separation 5-20 kpc, 20-50 kpc, 50-100 kpc, 100-150 kpc, respectively.
We note that the $m$ value decreases with increasing separation.
This result is further discussed in \S\ref{discussion}.

All these results show that the pair fractions do not change much 
with different $\Delta{z} = n\sigma_z$ criteria 
after applying the completeness and projection corrections, 
which indicates that our results are very robust;
however, the results do change with different projected separation criteria 
and absolute magnitude cuts.

We note that the role of photometric redshift in reducing the size
of the projection effect is relatively limited, 
as indicated by the similar error bars 
for the pair fractions in Figure~\ref{noback} for different $n$.
This is because the criterion of $\Delta m=\pm1$ effective eliminates
most of the foreground background galaxies.  
However, photometric redshift is essential in defining 
a volume-limited sample and deriving the redshift dependence of $N_c$.

\section{Discussions}\label{discussion}
\subsection{Major Merger Fraction}\label{merger_fraction}
Although we have applied projection correction for our pair analysis,
not all the close pairs we count will result in real mergers. 
Objects satisfying the pair criteria that are in the same structure 
(i.e., under each other's gravitational influence) 
are within 20 kpc projected distance, 
but not within 20 kpc in real 3-D space distance. 
\citet{YE1995} estimates that the true merger fraction ($f_{mg}$) is 
about half of the pair fraction which is evaluated
with a triple integral over projected and velocity separation 
by placing the correlation function into redshift space. 
Hence, we divide $N_c$ by 2 to calculate the merger fraction.

\subsubsection{Merger Timescale}\label{merger_timescale}
Close pairs are considered as early-stage mergers. 
To relate the close pairs to the overall importance of merger, 
the merger timescale ($T_{mg}$) has to be estimated. 
Following previous studies \citep{binney1987,patton2000},
assuming circular orbits and a dark matter density profile given 
by $\rho(r) \propto r^{-2}$,
the dynamical friction timescale ($T_{mg}$ in Gyr) is given by:
\begin{eqnarray}\label{timescale}
T_{mg} = \frac{2.64 \times 10^5r^2v_c}{M \ln \Lambda},
\end{eqnarray}
where $r$ is the physical pair separation in kpc, 
$v_c$ is the circular velocity in km s$^{-1}$, 
$M$ is the mass in $M_\odot$, and ln $\Lambda$ is the Coulomb logarithm. 
We adopt the average projected separation ($\sim$ 15 kpc)
as the physical pair separation since the major merger fraction
has been corrected from projected separation to three-dimensional separation.
We also assume $v_c \sim 250$ km s$^{-1}$. 
The mean absolute magnitude of companions is $M_{R_c} \sim -20.5$. 
If the mass-to-light ratio of $M/L \sim 5$ is assumed, 
we derive a mean mass of $M \sim 5\times10^{10} M_\odot$. 
\citet{dubinski1999} estimates ln $\Lambda \sim 2$.
With these values, we find $T_{mg} \sim 0.5$ Gyr. 
We note that this value is just a rough estimate over systems 
with a wide range of merger timescales,
but it still represents the average merger timescale in our sample.

\subsubsection{Merger Rate}\label{merger_r}
\begin{figure}
\plotone{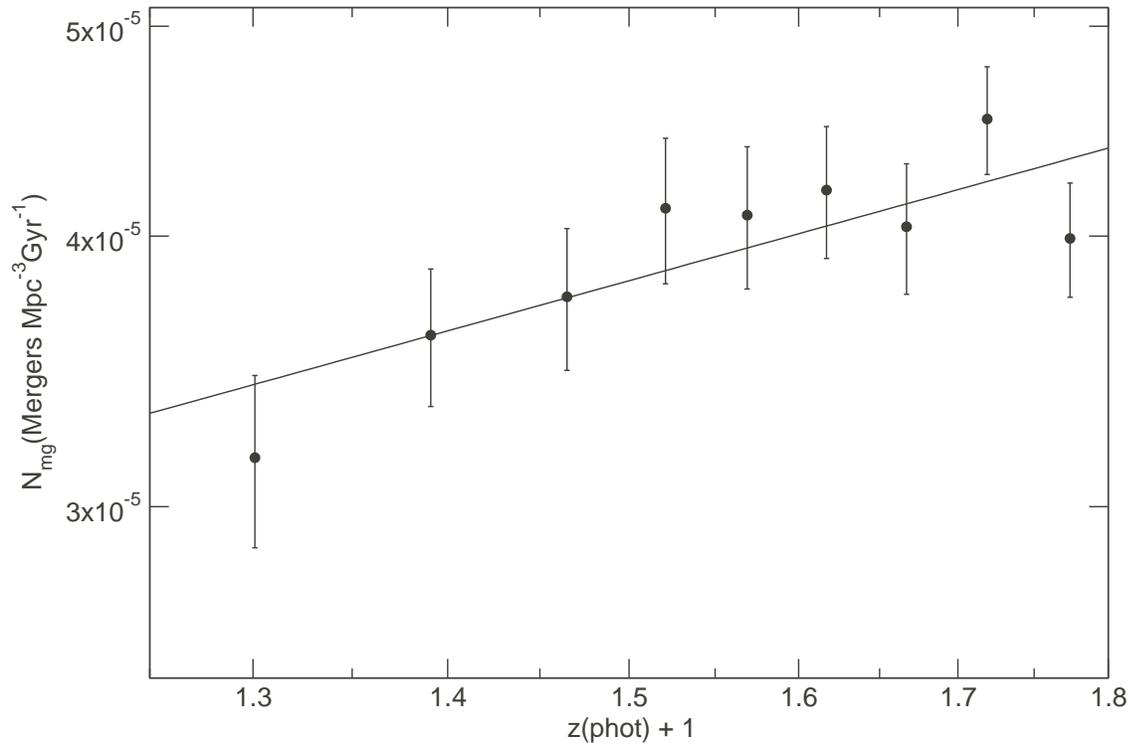}
\caption[Comoving merger rate vs. photometric redshift.]
{Comoving merger rate vs. photometric redshift. 
Each data point includes 17,500 objects 
and the error bars do not include the uncertainties in $T_{mg}$ and $C_{mg}$. 
The solid line is a fit of the form $(1+z)^\beta$, 
with the slope $\beta\sim0.8$.
\label{merger_rate} }
\end{figure}

The comoving merger rate is defined as 
the number of mergers per unit time per unit comoving volume. 
According to \citet{lin2004}, it can be estimated as:
\begin{eqnarray}\label{mg}
N_{mg} = 0.5n(z)N_c(z)C_{mg}T_{mg}^{-1},
\end{eqnarray}
where $T_{mg}$ is the dynamical friction timescale, 
$C_{mg}$ indicates the fraction of galaxies in close pairs 
that will merge in $T_{mg}$, 
and $n(z)$ is the comoving number density of galaxies. 
The factor 0.5 is to convert the number of galaxies 
into the number of merger events. 
We adopt $T_{mg} = 0.5$ Gyr and $C_{mg} = 0.5$, as discussed above.
We compute $n(z)$ using the primary sample, 
with the weight corrections included.  
We note that the computed $n(z)$ show a slight decline at $z>0.5$, 
which may be an indication of the effect of 
using a simple luminosity evolution description of $Qz$ for $M^*$ 
for choosing the sample.
The evolution of the merger rate is shown in Figure~\ref{merger_rate}. 
Each data point includes 17,500 objects 
and the error bars do not include the uncertainties in $T_{mg}$ and $C_{mg}$. 
Comparing our merger rate to \citet{lin2004} 
measured between $z\sim0.5$ to 1.2, our value is about 3 times lower. 
The primary reason is likely that we use $\Delta{R_c} \leq 1$ mag 
and search for major mergers with mass ratio from 1:1 to 3:1, 
but \citet{lin2004} look for mergers with mass ratio from 1:1 to 6:1. 
Our results show an increase in the merger rate with redshift of the form
$(1+z)^\beta$, with $\beta\sim0.8$.

\subsubsection{Major Merger Remnant Fraction}\label{merger_remnant}
With the derived major merger rates for several different epochs 
for $z < 0.8$ and the merger timescale, 
we can use these parameters to estimate the fraction of present day galaxies 
which have undergone major mergers in the past. 
These galaxies are merger remnants, 
and the fraction of the merger remnants is defined 
as the remnant fraction ($f_{rem}$). 
According to \citet{patton2000}, the remnant fraction is given by
\begin{eqnarray}\label{remnant_fraction}
f_{rem} = 1 - \prod^N_{j=1}\frac{1 - f_{mg}(z_j)}{1 - 0.5f_{mg}(z_j)}
\end{eqnarray}
where $f_{mg}$ is the merger fraction, 
$z_j$ is the redshift which corresponds to a lookback time of $t = jT_{mg}$, 
and j is an integer factor. 
Because the mass ratio of galaxies in a pair for this study is from 1:1 to 3:1, 
the merger remnant fraction is for major mergers. 
After applying this equation to our pair result, 
the estimated remnant fraction is $f_{rem} = 0.06$, 
which implies that $\sim 6\%$ of galaxies 
with $-25 \leq M_{R_c} \leq -20$ have undergone a major merger 
since $z \sim 0.8$.
 
\subsection{Evolution of the Pair Fraction}\label{interpretation}
The $m$ values determined in many previous observational results 
in similar redshift ranges are very diverse ($0 \leq m \leq 4$).
These results from the literature are listed in Table~\ref{m-value} 
in the order of redshift, and briefly discussed in \S\ref{introduction}. 
From $\Lambda$ CDM $N$-body simulations 
\citep{governato1999, gottlober2001}, 
the merger rates of halos increases with redshift as $(1+z)^m$, 
with $2.5 \leq m \leq 3.5$.
Our result for $-25 \leq M_{R_c} \leq -20$ is broadly consistent 
with the $N$-body simulations as well as 
all the previous observational results,
except those of \citet{lin2004} and \citet{bundy2004}.
Most of the previous works, except that of \citet{kartaltepe2007}, used 
spectroscopic redshifts to study pairs, 
especially for the lower redshift ranges.
Due to much smaller samples ( $< 1/20$ of our sample)
these studies need to convert a flux-limited sample to a volume-limited sample;
this conversion may affect the final results.
We note that all the previous studies did not exclude possible
cluster environments, which may cause a problem of transforming $N_c$ to
pair fraction, just because a galaxy would have a much higher chance to
have more than one companion in cluster environments, relative to in
field environments; the value of average number of companions per galaxy
is not a good representation of pair fraction anymore.
We do not have this problem because potential cluster members
in our sample are removed, and less than 2\% of pairs 
actually consists of triples or more in our work.

\begin{deluxetable}{lccc}
\tabletypesize{\tiny}
\tablecolumns{4}
\tablewidth{0pt}
\tablecaption{$m$ values from different references\label{m-value}}
\tablehead{
\colhead{Sample} & \colhead{$m$ value} & \colhead{Redshift Range} &
\colhead{Photometry Criteria}}
\startdata
This Work                               &  $2.83 \pm 0.33$ & 
0.25 $\leq z \leq$ 0.80 & -25 $\leq M_{R_c} \leq$ -20, field galaxies \\
\citet{governato1999, gottlober2001}    &  $2.5 \sim 3.5$ & &
$\Lambda$ CDM $N$-body simulations \\ 
\citet{zepf1989}                        &  $4.0 \pm 2.5$ & $z < 0.25$  &
$B \leq 22$ \\
\citet{YE1995}                          &  $4.0 \pm 1.5$ & $z < 0.38$  &
$r \leq 21.5$ \\
\citet{carlberg1994}                    &  $3.4 \pm 1.0$ & $z < 0.4$   &
$V \leq 22.5$ \\
\citet{patton1997}                      &  $2.8 \pm 0.9$ & $z \leq 0.45$ &
$r \leq 22$ \\
\citet{patton2002}                      &  $2.3 \pm 0.7$ & 
0.12 $\leq z \leq$ 0.55 & $R_c \leq 21.5$, Q=1 \\
\citet{burkey1994}                      &  $3.5 \pm 0.5$ & $z < 0.7$   &
$I \leq 22.3$  \\
\citet{lefevre2000}                     &  $2.7 \pm 0.6$ & $z < 1$     &
$I_{AB} \leq 22.5$ for primary, $I_{AB} \leq 24.5$ for secondary  \\
\citet{lin2004}                         &  $1.60 \pm 0.29$ & $0.45 < z < 1.2$ &
$R_{AB} \leq 24.1$ \\
\citet{lin2004}                         &  $0.51 \pm 0.28$ &  & Q=1 \\
\citet{kartaltepe2007}                  &  $3.1 \pm 0.1$ & $0 < z < 1.2$ &
$M_V < -19.8$, photometric redshift\\
\citet{kartaltepe2007}                  &  $2.2 \pm 0.1$ &  & Q=1 \\
\citet{bridge2007}                      &  $2.12 \pm 0.93$ & $0.2 < z < 1.3$ &
$ L_{IR} \leq 10^{11}L_{\odot}$ \\
\citet{bundy2004}                       &  no evolution & $0.5 < z < 1.5$ &
$K \leq 22.5$ \\
\citet{cassata2005}                     &  $2.2 \pm 0.3$ & $z < 2$     &
$K_s < 20$ \\
\enddata
\end{deluxetable}

The only previous work using a similar technique (photometric redshifts) 
with comparable size (59,221 galaxies, about 2.5 times smaller than our sample)
is \citet{kartaltepe2007}.
They have a higher redshift limit ($z = 1.2$) comparing with ours ($z = 0.8$),
while we have a considerably larger survey field 
(33.6 deg$^{2}$ vs. 2 deg$^{2}$).
They also have a similar luminosity cut 
($M_V = -19.8$, equivalent to $M_{R_c} \sim -20.3$) 
to that of our primary sample ($M_{R_c} = -20.0$),
although we have a one-magnitude deeper luminosity cut 
for the secondary sample ($M_{R_c} = -19.0$) 
which properly deals with the boundary problem (see \S\ref{boundary_effect}).
Since we apply luminosity evolution in our analysis,
we compare our result to the $m$ value 
of \citet{kartaltepe2007} derived with luminosity evolution.
The $m$ value of our study is somewhat higher,
$2.83 \pm 0.33$ vs. $2.2 \pm 0.1$, but within statistical consistency.

We further separate our sample with different luminosities, 
as shown in Figure~\ref{M_Rc}. 
In general, the pair fraction for luminous galaxies is lower than that of the 
fainter sample.
The $m$ values are $3.25 \pm 0.11$ and $1.79 \pm 0.53$
for $-21 \leq M_{R_c} \leq -20$ and $-25 \leq M_{R_c} < -21$, respectively.
From the result of the $m$ values, 
it appears that the pair fraction for the faint galaxy sample evolves
more rapidly than the luminous galaxy sample.
However, we note that the evolution of $N_c$ for the bright sample
may not fit a single power law very well.
The data suggest that $N_c$ may level out at $z \lesssim 0.6$,
while at $z \gtrsim 0.6$, the $m$ value is similar 
to that for the faint sample.
It is not clear what produces this possible leveling 
of the evolution for the bright sample.
A larger sample of galaxies will be useful in examining
the dependence of the pair fraction evolution as a function of
luminosity or stellar mass.

We also study the evolution of pair fractions 
for different projected separations. 
As shown in Figure~\ref{sep-1}, 
the $m$ value decreases with increasing separation;
i.e., the evolution gets weaker for larger separation.
This suggests that the timescale of a galaxy going from 20 kpc to 0
(i.e., merged), relative to the timescale for galaxies going in at 50 kpc,
increases at higher redshift, 
so that there is a build up of galaxies at 5-20 kpc.
Alternatively, this difference in $m$ values may be 
the steepening of the galaxy - galaxy correlation function
at very small scales at higher redshift.
Such steepening, at somewhat larger radius, has been observed
\citep{pollo2006,coil2006}.
The difference of timescale could be due to higher concentration
in the galaxy haloes at lower redshift.
For a given mass of dark halo, the density is higher in the inner region
for a high concentration halo, relative to a low concentration halo;
however, the density is lower in the outer region for a high concentration halo.
Therefore, for a high concentration halo, the dynamical friction timescale
would be longer in the outer region and shorter in the inner region.
Hence, our results suggest that for a given mass,
the galaxy dark halo size is smaller at lower redshift,
which is consistent with the simulation results from \citet{bullock2001}.
Bullock et al. studied dark matter halo density profile 
parameterized by an NFW \citep{nfw1997} form 
in a high-resolution $N$-body simulation of a $\Lambda$CDM cosmology.
Figure 10 in \citet{bullock2001} shows that the concentration parameter
increases with decreasing redshift for a given mass,
which may be responsible for producing the dependence of
the $m$ value evolution of radial bins.

\section{Conclusions}\label{conclusion}
We study the evolution of the pair fraction for over 157,000 galaxies in field 
with $0.25 \leq z \leq 0.8$, $\sigma_z/(1+z) \leq 0.3$, 
$-25 \leq M_{R_c} \leq -20$ for the primary sample,
$-26 \leq M_{R_c} \leq -19$ for the secondary sample, 
using 5 kpc $\leq d_{sep} \leq$ 20 kpc, $\Delta{z} \leq 2.5\sigma_z$, 
and $\Delta{R_c} \leq 1$ mag criteria 
from the RCS photometric redshift catalog. 
Our result for all the objects in the sample shows that 
the pair fraction increases with redshift as $(1 + z)^m$ 
with $m = 2.83 \pm 0.33$, which is consistent with $N$-body simulations
and many previous works.
We also estimate the major merger remnant fraction, which is 0.06.
This implies that only $\sim 6\%$ of galaxies
with $-25 \leq M_{R_c} \leq -20$ have undergone major mergers 
since $z \sim 0.8$.

By looking at the results separated into different magnitude bins, 
we find that the brighter the luminosity, the weaker the evolution.
We also study the evolution of pair fractions
for different projected separation bins
and find that the $m$ value decreases with increasing separation,
which suggests that for a given mass, 
the galaxy dark halo size is smaller at lower redshift,
which is consistent with the simulation results from \citet{bullock2001}.
In a future paper we will examine the evolution of the pair fraction 
in other environments (e.g., cluster core, cluster outskirts) 
to study whether the merger rate is affected by the environment.

\end{document}